\documentclass[twocolumn,floatfix,superscriptaddress,showpacs,showkeys]{revtex4}
\usepackage{graphicx}
\usepackage{amsmath}
\usepackage{booktabs}
\usepackage{amssymb}
\usepackage{multirow}
\usepackage{makecell}
\usepackage{textcomp}
\usepackage{subfigure}
\usepackage{phonetic}
\usepackage{extarrows}
\usepackage{color}
\usepackage{float}
\usepackage[colorlinks,citecolor=blue,linkcolor=blue]{hyperref}
\usepackage[utf8]{inputenc}
\begin{document}
\title{Signals from Fermionic inflationary cosmology with Yukawa interaction}
\author{Lin-Hong Sui}
\affiliation{College of Physics and Electronic Information, Inner Mongolia Normal University, 81 Zhaowuda Road, Hohhot, 010022, Inner Mongolia, China}
\affiliation{Inner Mongolia Key Laboratory of Applied Condensed Matter Physics, Inner Mongolia Normal University, 81 Zhaowuda Road, Hohhot, 010022, Inner Mongolia, China}
%\affiliation{School of Physics, Northeast Normal University, 5268 Renmin Street, Changchun 130024, China}
\author{Dan Li}
\affiliation{College of Physics and Electronic Information, Inner Mongolia Normal University, 81 Zhaowuda Road, Hohhot, 010022, Inner Mongolia, China}
\affiliation{Inner Mongolia Key Laboratory of Applied Condensed Matter Physics, Inner Mongolia Normal University, 81 Zhaowuda Road, Hohhot, 010022, Inner Mongolia, China}
\affiliation{School of Physical Science and Technology, Inner Mongolia University, No.235 West College Road, Hohhot, 010021, Inner Mongolia, China}
\author{Jia-Ze Sun}
\affiliation{College of Physics and Electronic Information, Inner Mongolia Normal University, 81 Zhaowuda Road, Hohhot, 010022, Inner Mongolia, China}
\affiliation{Inner Mongolia Key Laboratory of Applied Condensed Matter Physics, Inner Mongolia Normal University, 81 Zhaowuda Road, Hohhot, 010022, Inner Mongolia, China}
\author{Xi-Bin Li}
\email{lxbbnu@mail.bnu.edu.cn}
\affiliation{College of Physics and Electronic Information, Inner Mongolia Normal University, 81 Zhaowuda Road, Hohhot, 010022, Inner Mongolia, China}
\affiliation{Inner Mongolia Key Laboratory of Applied Condensed Matter Physics, Inner Mongolia Normal University, 81 Zhaowuda Road, Hohhot, 010022, Inner Mongolia, China}

\begin{abstract}
We investigate an inflationary model wherein the Dirac field $\psi$ is directly coupled to a scalar inflaton $\phi$ via a Yukawa interaction $g\phi\bar\psi\psi$ and examine the resulting observational implications.
Within the slow-roll approximation, we derive analytical solutions of the Dirac equations during inflation.
The analytical result on the fermion pair density $\langle n\rangle$ indicates that the Yukawa interaction strength $g$ is to characterize the degree of non-adiabaticity.
For large value of the dimensionless effective mass $\tilde m=(m+g\phi)/H$, i.e. $\tilde m\gtrsim 1$, the tensor-to-scalar ratio $r$ is suppressed by a factor of approximately $1/(1+2.95\pi^2g^2)$.
This condition is also characterized by a significant backreaction. Conversely, if $\tilde m \ll 1$, the value of $r$ remains consistent with that observed in standard cold inflation.
Our analysis is performed under the assumption of the highest inflationary energy scales compatible with current observational constraints.

%\pacs{42.50.Wk, 07.10.Cm, 42.50.Ar}
\end{abstract}
\maketitle

\section{\label{introduction}Introduction}

Gravitational particle production (as reviewed in Refs. \textcolor{blue}{\cite{production1,production2,production3}  }) is a generic phenomenon associated with quantum fields in a curved spacetime background.
In the context of Friedmann-Robertson-Walker cosmology devoid of inflationary processes, it has been established \textcolor{blue}{\cite{production4,production5}} that for conformally coupled scalar fields, fermions generation occurs close to the singularity of a radiation-dominated universe
when particle masses $m$ are comparable in magnitude to the Hubble expansion rate $H$.
Consequently, an upper bound of $10^9$ GeV is required on the mass of stable particles.
In contrast, in an inflationary universe, this bound would be removed since the singularity is replaced by a nearly constant Hubble parameter $H$ during the quasi-de Sitter era.
Hence, the potential existence of superheavy dark matter across a broad range of masses, including $m > H$, was established in Refs. \textcolor{blue}{\cite{production7,production8}}.
In these contexts, analytical formulas for the relic density have been derived in both the heavy and light mass regimes for conformally coupled scalars \textcolor{blue}{\cite{superheavy10,superheavy11}}.

The necessity to extend the Standard Model of particle physics is increasingly urgent, given the unresolved issues of baryogenesis and dark matter.
The Yukawa interaction, in which a Dirac field couples directly to a real scalar field, has been extensively studied as a simple dark matter candidate \textcolor{blue}{\cite{DM1,DM2,DM3,DM4}},
often referred to as "singlet fermionic dark matter". Phase transitions in this model have been thoroughly investigated \textcolor{blue}{\cite{Phase1,Phase2,Phase3}}.
Furthermore, numerous extensions of the Standard Model suggest the existence of superheavy dark matter candidates, which present compelling astrophysical implications.
For instance, Refs. \textcolor{blue}{\cite{reheating1,reheating2,reheating3,reheating4}}
focus on both numerical and analytical analyses of fermion production during the preheating phase.
Ref. \textcolor{blue}{\cite{production9}} indicates that when the inflaton is coupled to a massive particle, the resonant production of that particle during the inflationary period may generate sharp features within the primordial power spectrum.
The impact of radiative corrections on fermion dispersion relations and their association with particle production has also been analyzed \textcolor{blue}{\cite{production10,production11}}.

In this study, we investigate the gravitational particle production of Dirac fermions within the context of inflationary cosmology in the frame of Yukawa interaction.
As introduced previously, the masses of fermion can span a wide range, including light and heavy cases.
A key question is whether the (effective) fermion mass and the Yukawa coupling constant exert significant or distinctive impacts on cosmological observables, such as the tensor-to-scalar ratio.
Here, we derive the exact analytical solution of the Dirac equations within the quasi-de Sitter spacetime and provide more details about the observational effects including fermion density, modifications to the slow-roll parameters, and suppression of the tensor-to-scalar ratio.

This letter is organized as follows. In Sec. \ref{spectra}, we analytically solve the Dirac-like equation, incorporating Yukawa interactions in the inflationary universe.
In Sec. \ref{Obv}, we apply these results to analyze the inflationary cosmic parameters, including fermion pair density, slow-roll parameter, equation of state, and tensor-to-scalar ratio.
Finally, in Sec. \ref{conclusion}, a brief conclusion is provided.

\section{\label{spectra}Analytical solutions of the model}

The exponential expansion of the early universe is driven by a single scalar field, which can decay into various fundamental particles.
In this study, we explore the production of fermions during inflation via a Yukawa mechanism, where the fermions are directly coupled to the scalar field. The Lagrangian is given by \textcolor{blue}{\cite{Dirac1,Dirac2}}
\begin{align}
    \mathcal{L}=&\frac12\partial_\mu\phi\partial^\mu\phi-V(\phi)-g\phi\bar{\psi}\psi \nonumber\\
    &+\frac{\mathrm i}{2}[\bar{\psi} \bar{\gamma}^\mu D_\mu\psi-(D_\mu\bar{\psi})\bar{\gamma}^\mu\psi]-m\bar{\psi}\psi, \label{L}
\end{align}
where $\phi$ denotes the single scalar field responsible for cosmic inflation; $\psi$ and $\bar\psi=\psi^\dagger \gamma^0$ represent the Dirac spinor field and its adjoint,
respectively, coupled to the scalar field via the Yukawa interaction $-g\phi\bar{\psi}\psi$;
$V(\phi)$ is the potential of the scalar field; and $m$ is the static mass of the fermion.
The gamma matrices in Friedmann-Robertson-Walker universe are expressed as follows $\bar{\gamma}^0=\gamma^0,\quad \bar{\gamma}^i=a^{-1}\gamma^i$,
where $a$ is the cosmic scalar factor and $\gamma^\mu$ denotes the gamma matrices in flat spacetime, defined as
\begin{align}
    \gamma^0=\begin{pmatrix}I_{2\times2}&0\\0&-I_{2\times2}\end{pmatrix},\quad \gamma^i=\begin{pmatrix}0&\sigma^i\\-\sigma^i&0\end{pmatrix}. \label{gamma_mtrix}
\end{align}
Here $\sigma^i$ are the general Pauli matrices. The derivative operator in the Lagrangian \eqref{L} is expressed as \cite{Dirac1,Dirac2} $D_\mu\psi=\partial_\mu\psi+\Gamma_\mu\psi,\quad  D_\mu\bar{\psi}=\partial_\mu\bar\psi-\bar\psi\Gamma_\mu$,
with the spin connections in curved spacetime defined as $\Gamma_\mu$, given by $\Gamma_0=0,\ \Gamma_i=\frac{\dot{a}}{2}\gamma^0\gamma^i$.
The spatial contraction of these connections reads $\bar{\gamma}^i\Gamma_i=\frac32H\gamma^0$,
where $H$ represents the cosmic Hubble parameter.

The line element for the Friedmann-Robertson-Walker universe with tensor perturbation is given by
\begin{align}
    \mathrm{d}s^{2}&=\mathrm{d}t^2-a^2 \delta_{ij} \mathrm{d}x^i\mathrm{d}x^j \nonumber \\
    &=a^2{\left(\mathrm{d}\tau^2-\delta_{ij}\mathrm{d}x^i\mathrm{d}x^j\right)},  \label{metric}
\end{align}
with $\mathrm{d}t=a\mathrm{d}\tau $, where $t$ is the cosmic time and $\tau$ is the conformal time.
For exponentially expanding universe, the scalar factor can be expressed as $a=-1/H\tau$.

\subsection{\label{solution}Solution of Dirac-like equations}

The Dirac-like equation for the fermion field is obtained by varying the action with respect to the field $\psi$:
\begin{align}
    (\mathrm{i}\overline{\gamma}^\mu D_\mu-m-g\phi)\psi=0. \label{psi}
\end{align}
We now introduce the canonically normalized fermion field $u=\binom\varphi\eta$ in momentum space to rewrite the Dirac field as
$\psi=\int\mathrm d^3k\frac{\mathrm{e}^{\mathrm{i}\mathbf{k}\cdot\mathbf{x}}}{\left(2\pi a\right)^{3/2}}u $,
where $\varphi$ and $\eta$ are $2\times1$ undetermined matrices in momentum space.
Substituting this into the Dirac-like equation \eqref{psi} and using the contraction $\bar{\gamma}^i\Gamma_i=\frac32H\gamma^0$ to simplify it,
we obtain the following equations for $\varphi$ and $\eta$:
\begin{subequations}
\begin{align}
    &\mathrm{i}\frac{\partial}{\partial\tau}\varphi-(m+g\phi)a\varphi=k_i\sigma^i\eta, \label{varphi}\\
    &\mathrm{i}\frac{\partial}{\partial\tau}\eta+(m+g\phi)a\eta=k_i\sigma^i\varphi. \label{eta}
\end{align}
\end{subequations}
By multiplying $k_i\sigma^i$ on both sides of Eq. \eqref{varphi} and substituting it into Eq.\eqref{eta}, we derive the equation of motion:
\begin{align}
    \left[\frac{\partial^2}{\partial\tau^2}+k^2+(m+g\phi)^2a^2+\mathrm{i}g\phi^{\prime}a+\mathrm{i}(m+g\phi)a^{\prime}\right]\varphi=0. \label{Eq_a}
\end{align}
Here, the prime symbol $'$ denotes a partial derivative with respect to conformal time $\tau$.
For fermions traveling in a quasi-de-Sitter spacetime, scaled by $a=-1/H\tau$ ($H$ is slowly varying), the equation simplifies to
\begin{align}
    \left[\frac{\partial^2}{\partial\tau^2}+k^2+\frac{\left(m+g\phi\right)^2}{H^2\tau^2}+\mathrm{i}\frac{m+g\phi}{H\tau^2} %\right. \nonumber\\
    +\mathrm{i}\frac{g\dot{\phi}}{H^2\tau^2}+\mathrm{i}\frac{\dot{H}(m+g\phi)}{H^3\tau^2}\right]\varphi=0, \label{varphi_1}
\end{align}
where the dot $\dot{\ }$ represents the partial derivative to cosmic time $t$.

Using the slow-roll approximation, which assumes a quasi-exponential expansion during inflation, the following approximations hold:
\begin{align}
    \frac{g\dot{\phi}}{H^2},\ \frac{\dot{H}(m+g\phi)}{H^3}\ll\frac{m+g\phi}{H}. \label{approx}
\end{align}
Thus, the variables $\phi$, $\dot{\phi}$, $H$, and $\dot H$ can be treated as slowly varying parameters, similar to the general inflation model.
The solution to Eq. \eqref{varphi_1} is expressed in terms of Bessel functions, equivalent to the following equation:
\begin{align}
    y''+\frac{1-2\alpha}{z}y'+\left(\beta^2+\frac{\alpha^2-\mu^2}{z^2}\right)y=0, \label{EQ}
\end{align}
Correspondingly, the parameters are defined as $z=-k\tau$, $\alpha=1/2$ and $\beta=1$, and the order $\mu$ will be introduced in Eq. \eqref{mu}.
The general solution to Eq. \eqref{EQ} is a linear combination of Bessel functions with $y=z^\alpha Z_\mu(\beta z)$.
The Bunch-Davies vacuum provides the asymptotic flat form as the initial condition: $\varphi|_{\tau\to-\infty}\propto\mathrm e^{-\mathrm i k\tau}$.
Applying the asymptotic forms of the Bessel functions, the solution to \eqref{varphi_1} is
\begin{align}
    \varphi=C_1(-k\tau)^{1/2}\mathrm H_\mu^{(+)}(-k\tau)\,\chi_s, \label{varphi_2}
\end{align}
where $C_1$ is the undetermined coefficient, $\mathrm H_\mu^{(+)}$ is the Hankel function of the first kind, $\chi_s$ is the eigenvector of $\sigma_z$,
and $s=\pm$ denotes the spin state of a fermion. The order of the Hankel function is defined as follows:
\begin{align}
    \mu^2=\frac14-\tilde{m}^2-\mathrm{i}\tilde{m}-2\mathrm{i}\tilde{\delta}, \label{mu}
\end{align}
with the dimensionless parameters introduced as
\begin{align}
    \tilde{m}=\frac{m+g\phi}{H},\quad 2\tilde{\delta}=\frac{g\dot{\phi}+\dot{H}\tilde m}{H^2}.
\end{align}
The expression for $\mu$ involves two independent chiral indices:
\begin{align}
    \mu_h&=h\biggl[\left(-\frac12+\mathrm{i}\tilde{m}\right)^2-2\mathrm{i}\tilde{\delta}\biggr]^{1/2} \nonumber\\
    &=h\left(-\frac12+\tilde\eta+\mathrm i\tilde\xi\right), %\nonumber\\
    \label{mu_h}
\end{align}
where $h=\pm$ describes the particle or antiparticle state and the parameters are expressed to the first order as
\begin{subequations}
\begin{align}
    &\tilde\eta=\frac{\tilde\delta\tilde m}{\frac14+\tilde m^2}+\mathcal O\left(\tilde{\delta}^2\right), \\
    &\tilde\xi=\tilde m-\frac{\tilde\delta}{\frac12+2\tilde m^2}+\mathcal O\left(\tilde{\delta}^2\right).
\end{align}
\end{subequations}
In this paper, we take the order to be $\mu=\mu_+$; thus, we have $\mu_-=-\mu$.
The validity of approximation \eqref{mu_h} is based on the slow-roll approximation \eqref{approx}.
Therefore, the fermion state can be identified by chiral state $h$ and spin state $s$.

Consider the state $h=+,\ s=+$ as an example, where the spin eigenvector in solution \eqref{varphi_2} is $\chi_+=(1,\ 0)^T$.
Substituting this into Eq. \eqref{varphi}, we find the expression for the other undetermined matrix:
\begin{align}
    \eta=-C_1\frac{\mathrm i}{k}(-k\tau)^{1/2}\mathrm H_{\mu+1}^{(+)}(-k\tau)\binom{k_3}{k_+}, \label{eta_1}
\end{align}
where $k_\pm=k_x\pm\mathrm{i}k_y$. The expression for the Hankel function $\mathrm H_{\mu+1}^{(+)}$  is derived in detail in the appendix of Ref. \textcolor{blue}{\cite{chiral}}.
Thus, the solution for the fermion in the state $h=+,\ s=+$ is given by
\begin{align}
    u_{s=+}^{h=+}(\mathbf{k},\tau)
    = C_1\begin{pmatrix}\begin{pmatrix}1\\0\end{pmatrix}(-k\tau)^{1/2}\mathrm H^{(+)}_{-\frac12+\tilde\eta+\mathrm{i}\tilde{\xi}}(-k\tau)\\
    -\frac{\mathrm{i}}{k}\begin{pmatrix}{k_3}\\{k_+}\end{pmatrix}(-k\tau)^{1/2}\mathrm H^{(+)}_{\frac12+\tilde\eta+\mathrm{i}\tilde{\xi}}(-k\tau)\end{pmatrix},   \label{psi_1}
\end{align}
The normalization constant $C_1$ is determined by using the standard delta function normalization in the asymptotically flat-space limit \textcolor{blue}{\cite{Dirac3}}:
\begin{align}
    &(\psi_{\mathbf{k}}(\mathbf{x},\tau),\psi_{\mathbf{k}^{\prime}}(\mathbf{x},\tau)) \nonumber\\
    &=\int\mathrm{d}^3x\sqrt{\det g_{ij}}\psi_{\mathbf{k}}^\dagger(\mathbf{x},\tau)\psi_{\mathbf{k}^{\prime}}(\mathbf{x},\tau)\bigg|_{\tau\to-\infty} \nonumber\\
    &=\delta^3(\mathbf{k}-\mathbf{k}^\prime). \label{norm}
\end{align}
By applying the asymptotic form  of the Hankel function, $\mathrm H_\nu^{(+)}(z)\sim\sqrt{\frac2{\pi z}}\mathrm{e}^{\mathrm{i}(z-\frac12\nu\pi-\frac14\pi)}$ as $z\to+\infty$
and inserting \eqref{psi_1} into the normalization condition \eqref{norm}, we obtain the exact solution to the Dirac-like equation Eq. \eqref{psi} for the state $h=+$ and $s=+$:
%\begin{widetext}
\begin{subequations} \label{sol_psi}
\begin{align}
    u_{s=+}^{h=+}(\mathbf{k},\tau) = \frac1{\sqrt{2}}\Bigg(\frac{-\pi k\tau}{2\mathrm{e}^{\pi\tilde{\xi}}}\Bigg)^{1/2}
        \begin{pmatrix}\begin{pmatrix}1\\[0mm]0\end{pmatrix}\mathrm H^{(+)}_{-\frac12+\tilde\eta+\mathrm{i}\tilde{\xi}}(-k\tau)\\
    -\frac{\mathrm{i}}{k}\begin{pmatrix}{k_z}\\[0mm]{k_+}\end{pmatrix}\mathrm H^{(+)}_{\frac12+\tilde\eta+\mathrm{i}\tilde{\xi}}(-k\tau)\end{pmatrix}.
\end{align}
Similarly, we can derive the other three solutions:
\begin{align}
    u_{s=-}^{h=+}(\mathbf{k},\tau) = \frac1{\sqrt{2}}\Bigg(\frac{-\pi k\tau}{2\mathrm{e}^{\pi\tilde{\xi}}}\Bigg)^{1/2}
        \begin{pmatrix}\begin{pmatrix}0\\[0mm]1\end{pmatrix}\mathrm H^{(+)}_{-\frac12+\tilde\eta+\mathrm{i}\tilde{\xi}}(-k\tau)\\
    -\frac{\mathrm{i}}{k}\begin{pmatrix}{k_-}\\[0mm]{-k_z}\end{pmatrix}\mathrm H^{(+)}_{\frac12+\tilde\eta+\mathrm{i}\tilde{\xi}}(-k\tau)\end{pmatrix},
\end{align}
\begin{align}
    u_{s=+}^{h=-}(\mathbf{k},\tau) = \frac1{\sqrt{2}}\Bigg(\frac{-\pi k\tau}{2\mathrm{e}^{\pi\tilde{\xi}}}\Bigg)^{1/2}
        \begin{pmatrix}\begin{pmatrix}1\\[1mm]0\end{pmatrix}\mathrm H^{(+)}_{\frac12-\tilde\eta-\mathrm{i}\tilde{\xi}}(-k\tau)\\
    -\frac{\mathrm{i}}{k}\begin{pmatrix}{k_z}\\[1mm]{k_+}\end{pmatrix}\mathrm H^{(+)}_{-\frac12-\tilde\eta-\mathrm{i}\tilde{\xi}}(-k\tau)\end{pmatrix},  \\
    u_{s=-}^{h=+}(\mathbf{k},\tau) = \frac1{\sqrt{2}}\Bigg(\frac{-\pi k\tau}{2\mathrm{e}^{\pi\tilde{\xi}}}\Bigg)^{1/2}
        \begin{pmatrix}\begin{pmatrix}0\\[1mm]1\end{pmatrix}\mathrm H^{(+)}_{\frac12-\tilde\eta-\mathrm{i}\tilde{\xi}}(-k\tau)\\
    -\frac{\mathrm{i}}{k}\begin{pmatrix}{k_-}\\[1mm]{-k_z}\end{pmatrix}\mathrm H^{(+)}_{-\frac12-\tilde\eta-\mathrm{i}\tilde{\xi}}(-k\tau)\end{pmatrix}.
\end{align}
\end{subequations}

Equations \eqref{sol_psi} represent the solutions of the Dirac field equations in the inflationary universe.
The following analysis concerns their quantization, with the detailed calculations provided in the supplementary materials \textcolor{blue}{ \cite{SM} }.

\section{Observational effects \label{Obv}}
\subsection{Backreaction \label{backreaction}}

The equation of motion for the background inflaton, denoted as $\phi_0=\langle\phi\rangle$, is expressed by neglecting the second derivative with respect to cosmic time $t$,
\begin{subequations}
\begin{align}
    3H\dot{\phi}_0+V_\phi+g\left\langle\hat{\bar{\psi}}\hat{\psi}\right\rangle \approx 0.  \label{phi_0}
\end{align}
The term $V_\phi=\mathrm{d}V/\mathrm{d}\phi $ represents the derivative
of the potential $V$ with respect to the scalar field $\phi$.
To characterize the backreaction, we employ the slow-roll parameter $\varepsilon$.
Integrating the Friedmann equations for the Hubble parameter and its time derivative gives
\begin{align}
    &H^2\approx\frac1{3M_p^2}\left(V+\rho_D+\rho_Y\right),\\
    &\dot{H}\approx-\frac{\dot{\phi}^2}{2M_p^2}-\frac{\rho_\psi}{2M_p^2}-\frac{\rho_Y}{2M_p^2}-\frac{p_\psi}{2M_p^2},
\end{align}
\end{subequations}
where $p_\psi$ is the pressure of Dirac fermions. We then obtain the backreacting slow-roll parameter
\begin{align}
    \varepsilon=-\frac{\dot{H}}{H^2}
    =\frac{M_p^2}2\Bigg(\frac{V_\phi+g\left\langle\hat{\bar{\psi}}\hat{\psi}\right\rangle}
        {V+\rho_\psi+\rho_Y}\Bigg)^2+\frac32\frac{\rho_\psi+\rho_Y+p_\psi}{V+\rho_\psi+\rho_Y}. \label{back_epsilon}
\end{align}
In the above expression, the Yukawa energy density is defined as $\rho_Y=g\phi\left\langle\hat{\bar{\psi}}\hat{\psi}\right\rangle$.
In addition, the analytical expressions of the vacuum average of Dirac field and its adjoint $\left\langle\hat{\bar{\psi}}\hat{\psi}\right\rangle$, the energy density of Dirac field $\rho_\psi$, and the pressure $p_\psi$ are given by Eqs. \eqref{rho_psi} - \eqref{p_psi} in the supplementary materials \textcolor{blue}{\cite{SM}}.

\begin{figure}%[h]
	\center
	\includegraphics[width=\columnwidth]{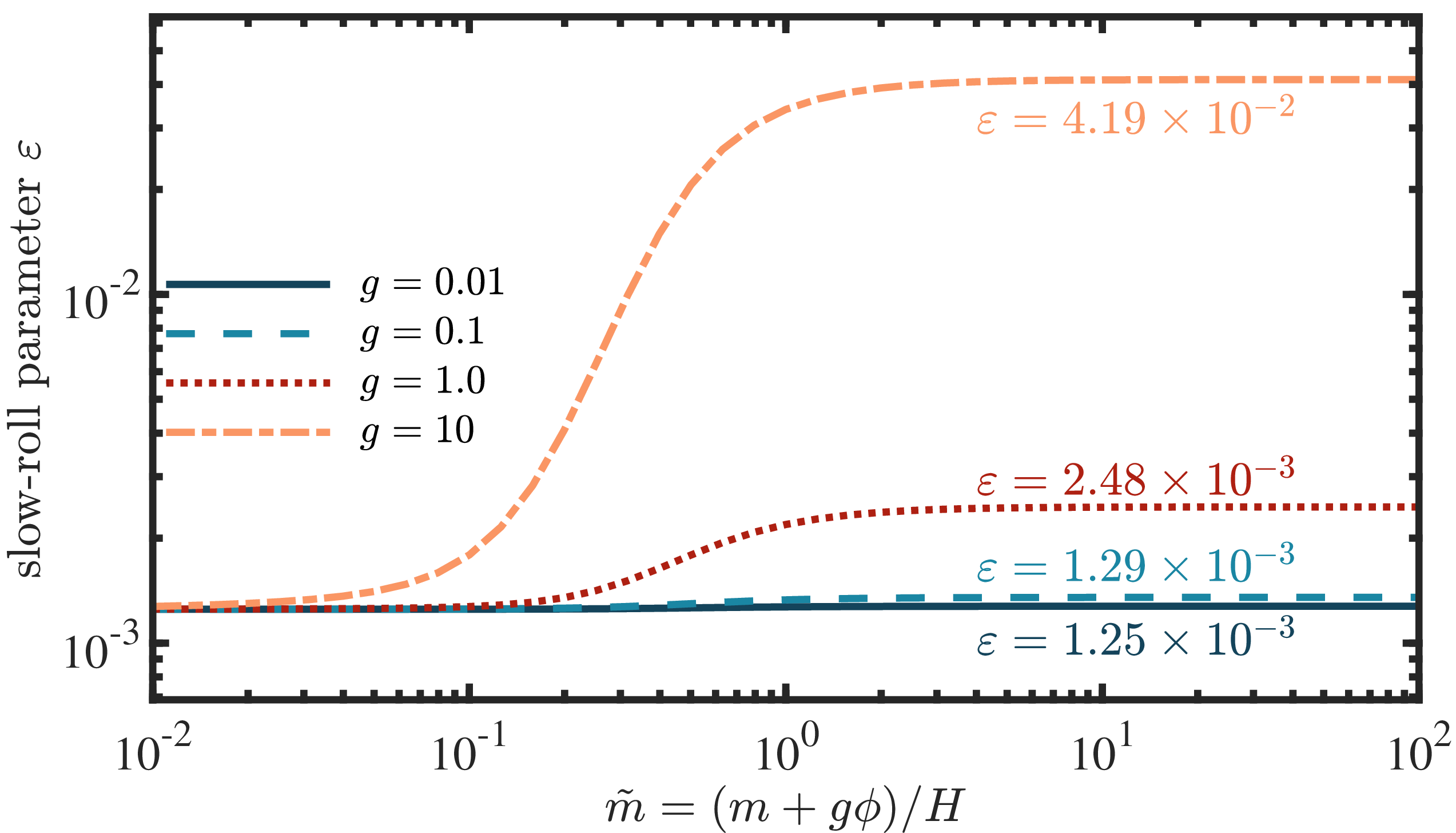}
	\caption{Slow-roll parameter $\varepsilon$ versus  $\tilde{m}$ for different values of the interaction strength g.}
	\label{epsilon_fig}
\end{figure}

To quantitatively illustrate the results, we set the parameters $V_\phi=10^{-13}M_p^3$ and $V=2\times10^{-12}M_p^4$ to ensure $H\sim10^{-5}M_p$, as constrained by the Planck 2018 data \textcolor{blue}{\cite{Planck}}.
The backreacting slow-roll parameters \eqref{back_epsilon} are presented as a function of the dimensionless effective mass $\tilde m$ for various values of $g$.
These curves are plotted in Fig. \ref{epsilon_fig}, where we have also applied the approximation $g\phi/H \approx \tilde m$.
The black solid quasi-horizontal line denotes the case with $g=0.01$, which is nearly equivalent to the general slow-roll parameter \(\varepsilon_0\).
The slow-roll parameter is evaluated by $\varepsilon_0 = ({M_p^2}/{2}) ({V_\phi}/{V})^2 = 1.25 \times 10^{-3}$, consisting of the most recent Planck + BICEP/Keck data with $r<0.036$ at 95\% confidence \textcolor{blue}{\cite{{Planck,BICEP}}}.
The direct interpretation of the figure suggests that a substantial backreaction effect only occurs when the effective mass is sufficiently large and the Yukawa coupling strength is notably strong.

\begin{figure}%[h]
	\center
	\includegraphics[width=\columnwidth]{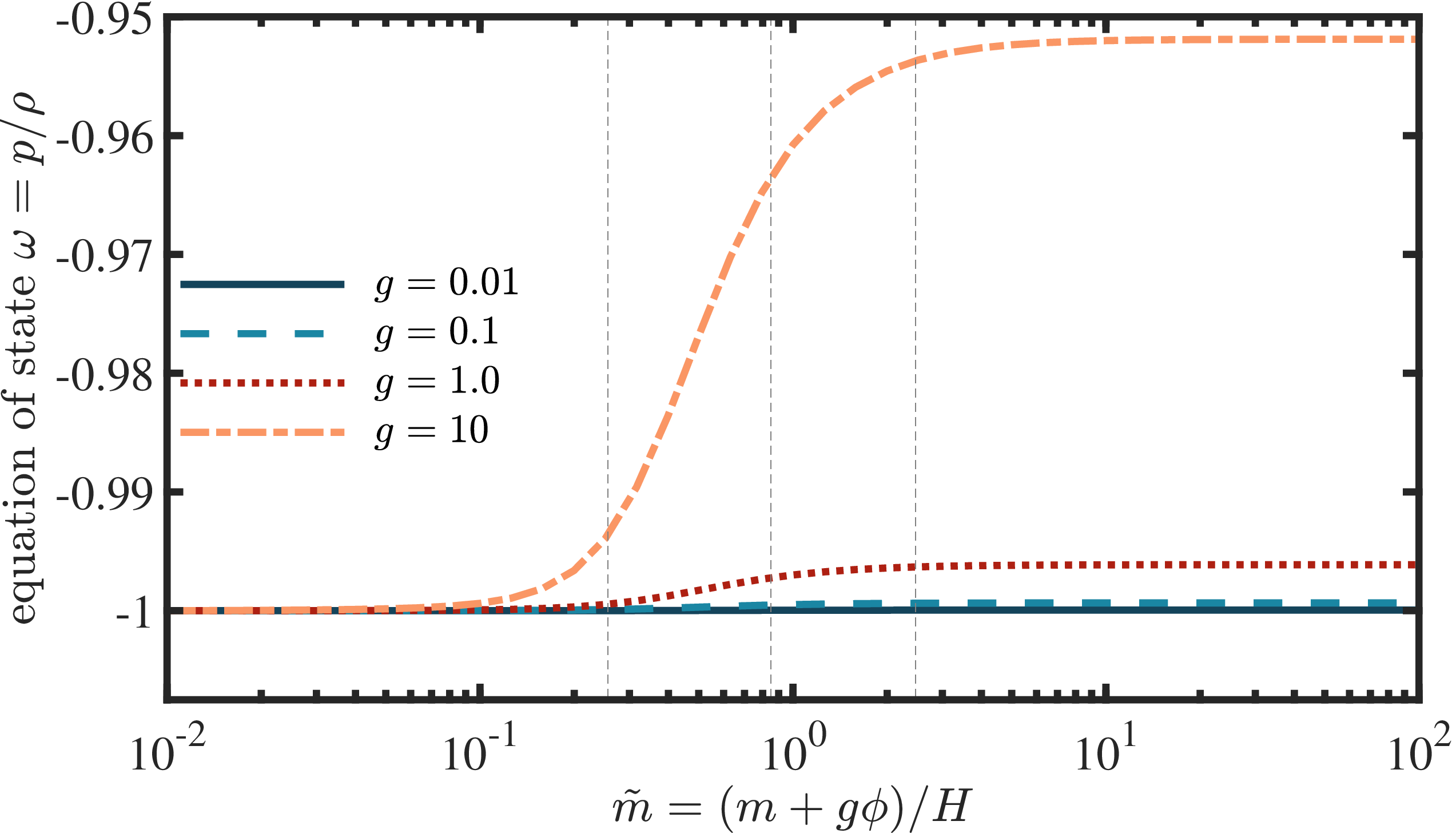}
	\caption{Equation of state parameter $\omega$ versus  $\tilde{m}$ for different values of the  interaction strength g.}
	\label{omega_fig}
\end{figure}

Notice that the Pauli exclusion principle for of Dirac fermions implies a positive pressure,
thereby influencing the equation of state $\omega=p/\rho$, with $p=-V+p_\psi$ and $\rho=V+\rho_\psi+\rho_Y$.
Figure \ref{omega_fig} shows the equation of state $\omega$ as functions of effective mass $\tilde m$ for various values of $g$.
We observe that only a strong interaction combined with a large effective mass can noticeably affect the exponential expansion; however, the resulting modification in amplitude remains relatively small.
This result indicates that $\omega$ is a model-dependent parameter.
For example, in the "Runaway" potential \textcolor{blue}{\cite{Runaway}} $V\propto\exp(-\alpha\phi^n)$, as the inflaton evolves, the impact on the equation of state becomes increasingly significant.
The three vertical dashed lines in Fig. \ref{omega_fig}, from left to right, correspond to the Dirac field equation-of-state parameters $\omega_\psi=p_\psi/\rho_\psi=$0.30, 0.15, and 0.03, respectively.
From this perspective, the backreation effect is closely related to the relativity (or degeneracy) of the Fermi gas.

\subsection{Fermion pair density \label{n_inf}}

We first analyze the chiral charge density produced from this model.
Chiral charge density, which represents the difference between the number densities of left-handed and right-handed particles,
can be presented as the time-component of chiral current $J_5^\mu=\langle\hat{\psi}^\dagger\gamma^0\gamma^\mu\gamma^5\hat{\psi} \rangle$, yielding a vanishing result %from Eqs. \eqref{sol_psi_J}
\begin{align}
    Q_5=\langle\hat{\psi}^\dagger\gamma^5\hat{\psi} \rangle=0. \label{q_inf}
\end{align}
This result arises from the absence of a source for parity violation. To achieve a non-vanishing chiral charge density, we should introduce sources that break parity symmetry, such as a magnetic field \textcolor{blue}{\cite{chiral}}, as well as Abelian and non-Abelian fermionic axions \textcolor{blue}{\cite{axion1,axion2}}, among others.
For instance, the non-trivial CP-violating vacuum structure of SU(2)-axion inflationary models naturally provides an efficient mechanism for generating massive fermions \textcolor{blue}{\cite{axion3}}.
Pseudo-scalar axion inflation offers a natural mechanism for generating significant non-Gaussianity in both single-field and multi-field slow-roll inflation \textcolor{blue}{\cite{axion4}}.
The resulting phenomenological signatures are highly distinctive, featuring significant non-Gaussianity of equilateral shape, along with notably large values of both the scalar spectral tilt and the tensor-to-scalar ratio.

The net number density, representing the number density of fermions minus the number density of anti-fermions, is calculated as (see the supplementary material \textcolor{blue}{\cite{SM}} for details)
\begin{align}
    \left\langle Q\right\rangle=\frac1{\mathrm{vol}(\mathbb{R}^3)}\int\mathrm d^{3}x\left\langle \hat{\bar\psi}(x)\gamma^0\hat\psi(x)\right\rangle=0,  \label{Q}
\end{align}
which is also straightforward to verify by considering the in vacuum state, i.e., $\left\langle Q\right\rangle|_{\tau\to-\infty} = 0$.
The vanishing result is due to the absence of a CP-violating source.
However, in the case of dark sectors that were never in thermal equilibrium with the hot primordial plasma, it is possible to produce equal numbers of fermions and anti-fermions.
Because they never thermalize, these particles do not annihilate efficiently.
As a result, while the net fermion number remains zero, the total energy density stored in the
fermions and anti-fermions can be substantial and non-zero.

The total density of fermions in comoving coordinates,
representing the number density of fermions plus the number density of anti-fermions, is defined as
\begin{align}
    \left\langle n\right\rangle=\frac1{\mathrm{vol}(\mathbb{R}^3)}\int\mathrm{d}^3x\left\langle:\hat{\psi}^\dagger(\mathbf x,\tau)\hat{\psi}(\mathbf x,\tau):\right\rangle. \label{n_x}
\end{align}
The details regarding the analytical expressions in the supplementary material \textcolor{blue}{\cite{SM}} imply that the adiabatic fermion pair density $\langle n_0\rangle$ is independent of the parameter $\tilde{\xi}$. This indicates insensitivity to both the strength of the Yukawa interaction and the inflationary models, while maintaining a constant value:
\begin{align}
    \langle n_0\rangle=0.2122(4\pi^2 H^3),  \label{n0}
\end{align}
which has also been numerically tested with minimal error.
This situation represents exact exponential expansion, i.e. $\phi=$constant, $\tilde\delta=\varepsilon=0$ and $\tilde\xi=\tilde m$=constant, interpreted as total adiabaticity.
However, in the general case, the slow-roll approximation implies that $\tilde\eta\neq0$,
which indicates non-adiabatic effects depend on the relevant inflationary parameters.
For example, in this case, $\langle n^\prime\rangle$ in Eq. \eqref{n} is expressed as the exact numerical result
\begin{align}
    \langle n^\prime\rangle
    =0.1414(4\pi^2 H^3)\tilde\eta \tanh\pi\tilde{\xi},   \label{n}
\end{align}
which is dependent of $\tilde m$ and $g$.

\textcolor{black}{
The total density $\langle n_0\rangle$ is defined for the exact de Sitter spacetime for which the Hubble parameter $H$ is a constant in time and space, i.e., $\langle n_0(t_1)\rangle-\langle n_0(t_2)\rangle=0$.
This quantity behaves like vacuum energy and, therefore, does not exhibit the characteristics of particle production.
As discussed in Ref. \textcolor{blue}{\cite{Parker}}, it has been shown that there is no gravitational particle production for massless fermions in an exact Friedmann-Robertson-Walker geometry. Consequently, this contribution is unobservable and instead acts as a vacuum term.}

\begin{figure}%[h]
	\center
	\includegraphics[width=\columnwidth]{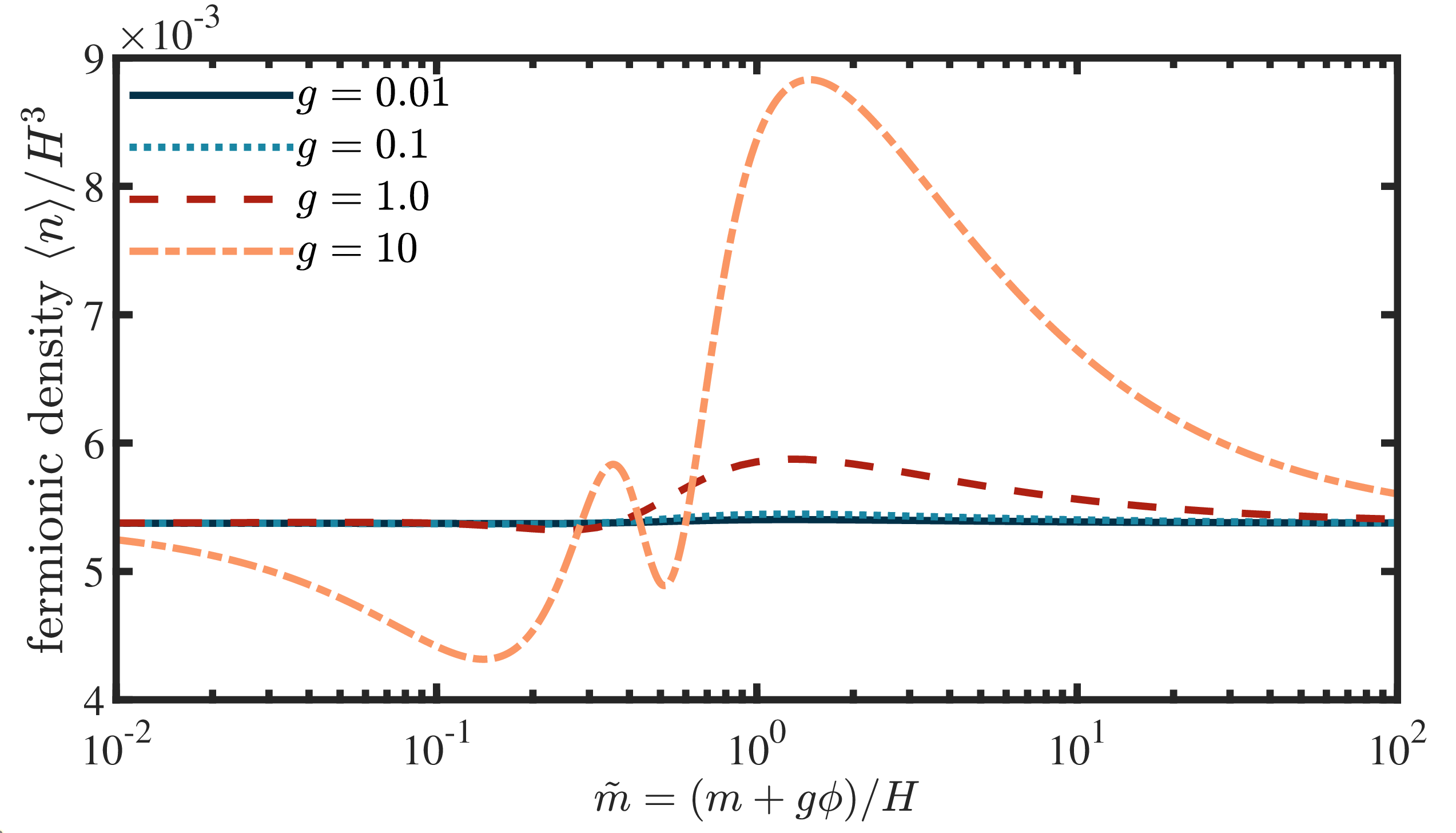}
	\caption{Comoving fermionic pair density $\langle n\rangle/H^3$ as a function of the
dimensionless effective mass $\tilde m$ with $g$=0.01, 0.1, 1, and 10, respectively.}
	\label{n_fig}
\end{figure}

Figure \ref{n_fig} shows the comoving fermionic pair density $\langle n\rangle$ as a function of the dimensionless effective mass $\tilde m$ with different values of Yukawa interaction strength $g$.
We have set $\varepsilon=-\dot H/H^2=1.25\times10^{-3}$ and $H/M_p=10^{-5}$ to ensure the amplitude of scalar power spectrum with $\mathcal P_\zeta=(1/8\pi^2\varepsilon)(H/M_p)^2\sim10^{-9}$.
It is observed that, in the case of $g\ll1$, the total fermion density is nearly equivalent to the result of $\langle n_0\rangle$, as expressed in Eq. \eqref{n0}.
As $g$ increases, the density exhibits a slight deviation from the average value $\langle n_0 \rangle$ in the vicinity of $\tilde{m} = 1$.
When the parameter $g$ increases to a sufficiently large value, it signifies a substantial oscillation around $\tilde m=1$.
We conclude that $\langle n_0\rangle$ with $\tilde{\eta} = 0$ represents the adiabatic case in which the fermionic comoving density remains constant as the universe undergoes completely exponential expansion.
When the Yukawa coupling constant is sufficiently weak, specifically when $g \ll 1$, $\langle n \rangle$ exhibits a near equivalence to the adiabatic case.
Non-adiabatic effects become significant when the coupling strength $g$ approaches or exceeds unity.
Therefore, the parameter $g$ can also serve as a descriptor for the degree of non-adiabaticity.
In addition, the first order of $\tilde{\eta}$ in Eq. \eqref{n3} consists of two contributions: the first is $-\ln4\cdot\langle n_0\rangle\tilde{\eta}\tanh\pi\tilde{\xi}-\langle n'\rangle$
and represents non-adiabaticity in the absence of oscillations, whereas the summation term accounts for the oscillatory component.

\subsection{Tensor-to-scalar ratio}

The scalar field driving cosmic inflation can be decomposed into its background component $\phi_0$ and its perturbation component $\delta\phi(\mathbf x,t)$, i.e., $\phi=\phi_0+\delta\phi$.
By varying the action Eq.\eqref{L} with respect to $\phi$ and subtracting the background field $\phi_0$, we obtain the equation of motion for the perturbed field
\begin{align}
    \left(\frac{\partial^2}{\partial t^2}+3H\frac\partial{\partial t}-\frac{\nabla^2}{a^2}+V_{\phi\phi}\right)\delta\phi=g({\bar{\psi}} {\psi}-\langle{\bar{\psi}} {\psi}\rangle), \label{delta_phi_x}
\end{align}
with $V_{\phi\phi}=\mathrm d^2V/\mathrm d\phi^2$.
This equation confirms the vanishing of the vacuum average $\langle\delta\phi\rangle=0$.
The equation of motion for tensor perturbation is derived from the Einstein equation and can be written as the energy-momentum tensor for the fermion field $T^{mn}$:
\begin{align}
    \left(\frac{\partial^2}{\partial\tau^2}+k^2-\frac{a^{\prime\prime}}a\right)(ah_{ij})=\frac{2a^3}{M_p^2}\Pi_{ij,mn}T^{mn}(\mathbf{k},\tau),
\end{align}
where $M_p=(8\pi G)^{-1/2}$ and the transverse-traceless operator $\Pi_{ij,mn}$ is defined in the supplementary materials.

The spectral densities of scalar perturbations $\mathcal P_\mathcal{R}$ and tensor perturbations $\mathcal P_h$ are defined as
\begin{align}
    &\left\langle\hat{\mathcal{R}}(\mathbf{k})\hat{\mathcal{R}}^\dagger(\mathbf{k}^{\prime})\right\rangle=\frac{2\pi^2}{k^3}\delta^3(\mathbf{k}-\mathbf{k}^{\prime})\mathcal{P}_{\mathcal{R}}(k),\\
    &\sum_{\lambda=\pm}\left\langle\hat{h}_\lambda(\mathbf{k})\hat{h}_\lambda^\dagger(\mathbf{k}^{\prime})\right\rangle=\frac{2\pi^2}{k^3}\delta^3(\mathbf{k}-\mathbf{k}^{\prime})\mathcal{P}_h(k).
\end{align}
As shown in Eqs. \eqref{delta_phi_k_1} and \eqref{h}, both perturbations have two contributions: one from vacuum and the other from the sourced Dirac field.
Since these two contributions are statistically independent, the total spectral densities can be written as $\mathcal{P}_{\mathcal{R}}=\mathcal{P}_{\mathcal{R}}^{(v)}+\mathcal{P}_{\mathcal{R}}^{(s)}$
and $\mathcal{P}_h=\mathcal{P}_h^{(v)}+\mathcal{P}_h^{(s)}$, with the standard vacuum spectra
\begin{align}
    \mathcal{P}_{\mathcal{R}}^{(v)}=\left(\frac{H^2}{2\pi\dot{\phi}_0}\right)^2,\quad \mathcal{P}_h^{(v)}=\frac{2H^2}{\pi^2M_p^2}.
\end{align}
The sourced spectra $\mathcal{P}_{\mathcal{R}}^{(s)}$ and $\mathcal{P}_h^{(s)}$ are obtained directly from Eqs. \eqref{corr_R} and \eqref{corr_h}.
Finally, the tensor-to-scalar ratio is expressed as
\begin{align}
    r=\frac{\mathcal{P}_h^{(v)}+\mathcal{P}_h^{(s)}}{\mathcal{P}_{\mathcal{R}}^{(v)}+\mathcal{P}_{\mathcal{R}}^{(s)}}. \label{r}
\end{align}

\begin{figure}%[h]
	\center
	\includegraphics[width=\columnwidth]{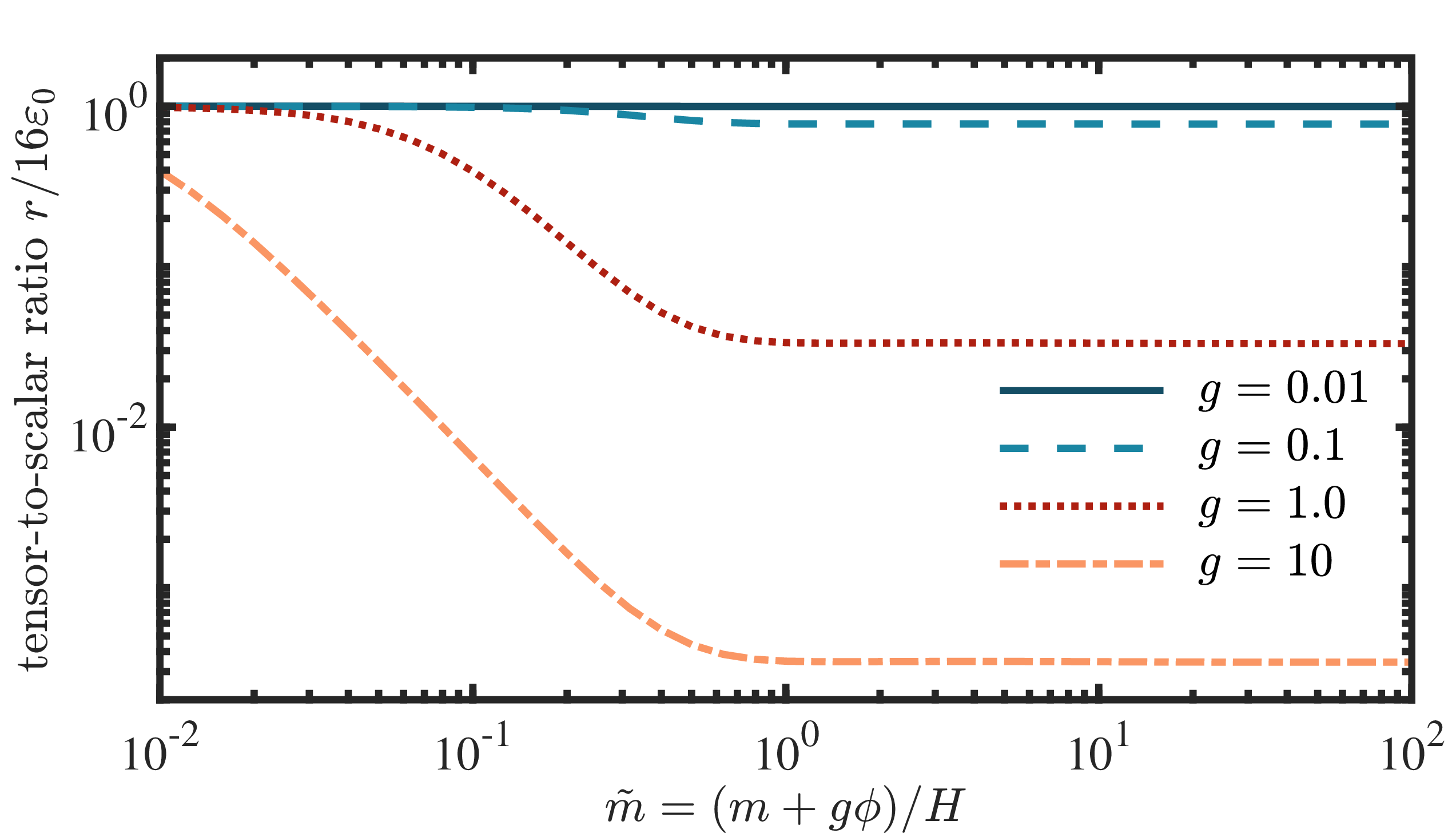}
	\caption{Tensor-to-scalar ratio $r/16\varepsilon_0$ as a function of dimensionless effective mass $\tilde m$ for different values of Yukawa coupling constant $g$, where $\varepsilon_0=1.25\times10^{-3}$. }
	\label{r_fig}
\end{figure}

Figure \ref{r_fig} plots the tensor-to-scalar ratio $r/16\varepsilon_0$ as a function of dimensionless effective mass $\tilde m=(m+g\phi)/H$,
where $\varepsilon_0$ is the general slow-roll parameter, already defined in Sec. \ref{backreaction}.
This parameter is set to $\varepsilon_0 = 1.25 \times 10^{-3}$ to incorporate the most recent data with $r<0.036$ at 95\% confidence \textcolor{blue}{\cite{BICEP}}.
It is evident that for sufficiently small Yukawa coupling $g$, the ratio $r$ closely follows that of standard single-field inflation.
However, as $g$ increases, the behavior can be described in three stages.
For small $\tilde m$, the ratio $r$ remains close to $16\varepsilon_0$, consistent with single-field inflation.
As $\tilde m$ increases, the ratio $r$ decreases gradually until it reaches a threshold.
At sufficiently large $\tilde m\gtrsim 1$, the ratio $r$ stabilizes,
with its amplitude compressed by approximately $1/(1+2.95\pi^2g^2)$.

\section{\label{conclusion}Conclusions}    %V

In this study, we investigated an inflationary model in which the scalar inflaton couples directly to the Dirac field through Yukawa interactions.
We theoretically calculated the fermion production rates during inflation.
In an exactly exponential expanding universe, corresponding to a de Sitter spacetime ($\varepsilon_0=0$), the total comoving fermion pair density is found to be constant with a value of $\langle n_0\rangle=0.2122(4\pi^2 H^3)$,
representing the exact adiabatic condition and demonstrating independence from both the effective mass $\tilde m=(m+g\phi)/H$ and the Yukawa interaction strength $g$.
Here, $H$ represents the inflationary Hubble parameter.
This density may be interpreted as a constant vacuum contribution rather than particle production.
However, under the slow-roll approximation ($\varepsilon_0\neq0$),
the fermionic pair density depends on the parameter $\tilde m$ and $g$.
For a light mass condition $\tilde m\ll1$, the fermion pair density stabilizes almost at the adiabatic case $\langle n_0\rangle$,
indicating an independence of the effective mass $\tilde m$ or the strength of the Yukawa interaction $g$.
As $g$ approaches unity, the density exhibits pronounced oscillations around $\tilde m\sim1$.
This result indicates that the Yukawa interaction strength $g$ characterizes the degree of non-adiabaticity during the inflationary phase.

By treating the inflationary era as a de Sitter spacetime, we analytically calculated the scalar and tensor perturbations sourced from the Dirac field.
When combined with the vacuum contributions, we derived the tensor-to-scalar ratio $r$.
For sufficiently small values of the dimensionless effective mass $\tilde m$ or Yukawa coupling constant $g$, tensor-to-scalar $r$ is equivalent to $16\varepsilon_0$, consistent with standard single-field cold inflation.
As $\tilde m$ increases, $r$ decreases smoothly until it reaches a constant value, with an amplitude compressed by a factor of $1/(1+2.95\pi^2g^2)$ when $\tilde m\gtrsim 1$.
Finally, we emphasize that these results are based on the assumption of the highest inflationary energy scales, which remain compatible with current data.

\section*{Acknowledgements}
This work was supported by the National Natural Science Foundation of China (Grant No. 12275143), Central Guidance for Local Science and Technology Development Fund Project (Grand No. 2024ZY0113, 2025ZY0020),
Inner Mongolia Natural Science Foundation (Grant No. 2021LHBS01001, 2024SHZR0009).
% and Scientific Research Funding Project for Introduced High Level Talents of IMNU (Grant No. 2020YJRC001).

%%\end{thebibliography}

\begin{widetext}

\section{Quantization of Dirac fields  \label{quan}}

Aside from the solutions expressed in terms of Hankel functions,
one may also use Bessel functions of the first kind $\mathrm J_{\pm \mu}$ to describe the propagation, as detailed in Sec. \ref{Bessel}.
In Eq. \eqref{sol_psi_J}, the eigenvectors $\tilde u$ represent particles propagating to the asymptotic future (particle states),
while eigenvectors $\tilde v$ represent particles propagating to the asymptotic past (antiparticle states).
Based on the transformation between Bessel functions and Whittaker functions,
the eigenvectors in terms of Hankel function (M18), which corresponds to the equation with the same number in the main text, can be expanded as a sum of eigenvectors \eqref{sol_psi_J} shown in the current document via a Bogoliubov transformation \textcolor{blue}{\cite{chiral}}
\begin{align}
    u^{h=\pm}_s=\alpha^\pm \tilde u^{h=\pm}_s + \beta^\mp \tilde v^{h=\pm}_s, \label{B_transf}
\end{align}
where $\tilde v^h_s=\tilde u^{-h}_s$ and the Bogoliubov coefficients
\begin{align}
    |\beta^\pm|^2=|\alpha^\mp|^2=\frac{\mathrm e^{\pm\pi\tilde\xi}}{2\cosh\pi\tilde\xi} \label{beta}
\end{align}
are independent of momentum $k$ and conformal time $\tau$.

The mode expansions for the quantized Dirac field and anti-Dirac field about the in-vacuum state are constructed as
\begin{subequations}\label{quantized_1}
\begin{align}
    &\hat{\psi}(\mathbf{x},\tau)\Big|_{\tau\to-\infty}=\sum_{s=\pm}\int\frac{\mathrm{d}^3k}{\left(2\pi a\right)^{3/2}}
    \left[\mathrm{e}^{\mathrm{i}\mathbf{k}\cdot\mathbf{x}}\hat{b}_s(\mathbf{k})u_s^{h=+}(\mathbf{k},\tau) % \right. \nonumber\\
    +\mathrm{e}^{-\mathrm{i}\mathbf{k}\cdot\mathbf{x}}\hat{d}_s^\dagger(\mathbf{k})u_s^{h=-}(\mathbf{k},\tau)\right], \\
    &\hat{\bar\psi}(\mathbf{x},\tau)\Big|_{\tau\to-\infty}=\sum_{s=\pm}\int\frac{\mathrm{d}^3k}{\left(2\pi a\right)^{3/2}}
    \left[\mathrm{e}^{-\mathrm{i}\mathbf{k}\cdot\mathbf{x}}\hat{b}_s^\dagger(\mathbf{k})u_s^{+\,\dagger}(\mathbf{k},\tau) %\right. \nonumber\\
    +\mathrm{e}^{\mathrm{i}\mathbf{k}\cdot\mathbf{x}}\hat{d}_s(\mathbf{k})u_s^{-\,\dagger}(\mathbf{k},\tau)\right]\gamma_0,    %\label{quantized_1}
\end{align}
\end{subequations}
where $\hat{b}_s(\mathbf{k})$ and $\hat{d}_s^\dagger(\mathbf{k})$ are the initial annihilation and creation operators for particles and antiparticles, respectively, with spin state $s$ and momentum $\mathbf k$.
The anti-commutation relations are given by
\begin{align}
    \left\{\hat{b}_s(\mathbf{k}),\hat{b}_{s^{\prime}}^\dagger(\mathbf{k}^{\prime})\right\}=\left\{\hat{d}_s(\mathbf{k}),\hat{d}_{s^{\prime}}^\dagger(\mathbf{k}^{\prime})\right\} %\nonumber\\
    =(2\pi)^3\delta^3(\mathbf{k}-\mathbf{k}^{\prime})\delta_{ss^{\prime}} \label{commu}
\end{align}
and $\{\text{otherwise}\}=0$. Inserting the Bogoliubov transformations \eqref{B_transf} into the quantization formulas \eqref{quantized_1},
the quantized Dirac field and anti-Dirac field at finite time become
\begin{subequations}\label{quantized_2}
\begin{align}
    &\hat{\psi}(\mathbf{x},\tau)=\sum_{s=\pm}\int\frac{\mathrm{d}^3k}{\left(2\pi a\right)^{3/2}}
    \left[\mathrm{e}^{\mathrm{i}\mathbf{k}\cdot\mathbf{x}}\hat{B}_s(\mathbf{k})\tilde{u}_s^{h=+}(\mathbf{k},\tau) %\right. \nonumber\\
    +\mathrm{e}^{-\mathrm{i}\mathbf{k}\cdot\mathbf{x}}\hat{D}_s^\dagger(\mathbf{k})\tilde{u}_s^{h=-}(\mathbf{k},\tau)\right], \\
    &\hat{\bar\psi}(\mathbf{x},\tau)=\sum_{s=\pm}\int\frac{\mathrm{d}^3k}{\left(2\pi a\right)^{3/2}}
    \left[\mathrm{e}^{-\mathrm{i}\mathbf{k}\cdot\mathbf{x}}\hat{B}_s^\dagger(\mathbf{k})\tilde{u}_s^{+\,\dagger}(\mathbf{k},\tau) %\right. \nonumber\\
    +\mathrm{e}^{\mathrm{i}\mathbf{k}\cdot\mathbf{x}}\hat{D}_s(\mathbf{k})\tilde{u}_s^{-\,\dagger}(\mathbf{k},\tau)\right]\gamma_0,
\end{align}
\end{subequations}
where the new creation/annihilation operators are related to the initial ones by
\begin{subequations}
\begin{align}
    \hat{B}_s(\mathbf{k})=\alpha^+\hat{b}_s(\mathbf{k})-\beta^+\hat{d}_s^\dagger(-\mathbf{k}),\\
    \hat{D}_s^\dagger(\mathbf{k})=\beta^-\hat{b}_s(\mathbf{k})+\alpha^-\hat{d}_s^\dagger(-\mathbf{k}).
\end{align}
\end{subequations}
The new operator $\hat B_s^\dagger$ represents a fermionic particle propagating to the asymptotic future in two ways: either as article from the in-vacuum state with amplitude $\alpha^+$ or through a transition from an antiparticle into a particle with amplitude $\beta^+$.
The same interpretation applies to the creation operators $\hat D_s^\dagger$,
meaning that particle/antiparticle production rates in the out-vacuum state are nonzero for
\begin{subequations} \label{average}
\begin{align}
    \left\langle\hat{B}_{s}^{\dagger}(\mathbf{k}) \hat{B}_{s^{\prime}}(\mathbf{k}^{\prime})\right\rangle=(2\pi)^{3}\delta_{ss^{\prime}}\delta^{3}(\mathbf{k}-\mathbf{k}^{\prime})\left|\beta^{+}\right|^{2},\\
    \left\langle\hat{D}_{s}^{\dagger}(\mathbf{k}) \hat{D}_{s^{\prime}}(\mathbf{k}^{\prime})\right\rangle=(2\pi)^{3}\delta_{ss^{\prime}}\delta^{3}(\mathbf{k}-\mathbf{k}^{\prime})\left|\beta^{-}\right|^{2},
\end{align}
\end{subequations}
and $\langle\text{otherwise}\rangle=0$, where $\langle\cdots\rangle$ represents the vacuum average.

\section{Hamiltonian and equation of state \label{H_omega}}
The Hamiltonian is expressed as
\begin{align}
    \hat H &= \int \mathrm d^3x\, \mathcal{H} =
    \int \mathrm d^3x\left[\mathrm i\hat \psi^\dagger(x)\partial_0\hat \psi(x)-\hat{\bar{\psi}}(x)(\mathrm i\bar \gamma^\mu\partial_\mu-m-g\phi)\hat \psi(x)\right]  \nonumber\\
    &=\int\frac{\mathrm d^3k}{(2\pi a)^3} \sqrt{(k/a)^2+(m+g\phi)^2}\sum_{s=\pm}
    \left[\hat{B}^\dagger_s(\mathbf{k})\hat{B}_s(\mathbf{k})\tilde{u}_s^{h=+\ \dagger}(\mathbf{k},\tau)\tilde{u}_s^{h=+}(\mathbf{k},\tau)
    -\hat{D}_s(\mathbf{k})\hat{D}^\dagger_s(\mathbf{k})\tilde{u}_s^{h=-\ \dagger}(\mathbf{k},\tau)\tilde{u}_s^{h=-}(\mathbf{k},\tau) \right]  \nonumber\\
    &=\sum_{s=\pm}\int\frac{\mathrm d^3k}{(2\pi a)^3} \sqrt{(k/a)^2+(m+g\phi)^2}
    \left[\hat{B}^\dagger_s(\mathbf{k})\hat{B}_s(\mathbf{k})\tilde{u}_s^{h=+\ \dagger}(\mathbf{k},\tau)\tilde{u}_s^{h=+}(\mathbf{k},\tau)
    +\hat{D}^\dagger_s(\mathbf{k})\hat{D}_s(\mathbf{k})\tilde{u}_s^{h=-\ \dagger}(\mathbf{k},\tau)\tilde{u}_s^{h=-}(\mathbf{k},\tau) \right]  \nonumber\\
    &\qquad\qquad\qquad\qquad\qquad\qquad\qquad\qquad\qquad\qquad
    -\frac{2}{a^3}\int\mathrm d^3k\, \delta^3(\mathbf 0)\sqrt{(k/a)^2+(m+g\phi)^2}\
    \tilde{u}_-^{-\ \dagger}(\mathbf{k},\tau)\tilde{u}_-^{-}(\mathbf{k},\tau).
\end{align}
with the applications of $\int \mathrm d^3x\, \mathrm e^{\pm\mathrm  i(\mathbf{k-p})\cdot\mathbf{x}}=(2\pi)^3\delta^{(3)}(\mathbf{k-p})$ and $\tilde{u}_s^{h\ \dagger}\tilde{u}_r^{h}\propto\delta_{sr}$.
Here, we have neglected the terms containing $\hat B^\dagger\hat D$ and $\hat B \hat D^\dagger$ since they vanish by vacuum average, and have used the anti-commutation relation \eqref{commu}.
Clearly, the last term in the last step is an infinite constant.
It corresponds to the sum over all the zero-point energies of the infinite eigenfunctions,
each with a contribution $\sqrt{(k/a)^2+\tilde m^2}\ \tilde{u}_s^{-\ \dagger}(\mathbf{k},\tau)\tilde{u}_s^{-}(\mathbf{k},\tau)$.
This phenomenon will always be presented in quantum field theory, much like the zero-point energy presented in the harmonic oscillator, and we will address it throughout the following courses.
However, since it is a constant, we can always shift the energy origin to cancel it.

The energy density of Dirac field is the vacuum average of the Hamiltonian density, given by
\begin{align} \label{rho_psi}
    \rho_\psi=\left\langle\mathcal H\right\rangle %\nonumber\\
    &=\frac{4\pi^2 H^4}{\cosh\pi\tilde{m}}\int_0^1\mathrm{d}z\, z^3\sqrt{z^2+\tilde m ^2}
     \cdot\left[\left|\mathrm J_{\frac12+\mathrm{i}\tilde{m}}\left(z\right)\right|^2
        +\left|\mathrm J_{-\frac12-\mathrm{i}\tilde{m}}\left(z\right)\right|^2\right] \nonumber\\
    &\qquad\qquad\qquad
    -\frac{2H^4}{a^3}\int\mathrm d^3k\, \delta^3(\mathbf 0)z^3\sqrt{z^2+\tilde m ^2}\
    \left[\left|\mathrm J_{-\frac12-\mathrm{i}\tilde{m}}\left(-k\tau\right)\right|^2+
    \left|\mathrm J_{\frac12+\mathrm{i}\tilde{m}}\left(-k\tau\right)\right|^2 \right],
\end{align}
with $z=-k\tau$.
Based on our analysis, the infinite term $\delta^3(\mathbf 0)$ will be disregarded.
Therefore, the analytical expressions for the vacuum expectation value of the Dirac field and its adjoint, $\left\langle\hat{\bar{\psi}}\hat{\psi}\right\rangle$, and the pressure of the Dirac field, $p_\psi$, are expressed as follows:
\begin{align}
    \left\langle\hat{\bar{\psi}}\hat{\psi}\right\rangle&=\pi\frac{|\beta^+|^2+|\beta^-|^2}{\cosh\pi\tilde{m}}\int\frac{\mathrm d^3 k}{a^3}(-k\tau)
    \cdot\left[\left|\mathrm J_{-\frac12-\mathrm{i}\tilde{m}}\left(-k\tau\right)\right|^2-\left|\mathrm J_{\frac12+\mathrm{i}\tilde{m}}\left(-k\tau\right)\right|^2 \right] \nonumber\\
    &=\frac{4\pi^2 H^3}{\cosh\pi\tilde{m}}\int_0^1\mathrm{d}z\, z^3
    \cdot\left[\left|\mathrm J_{-\frac12-\mathrm{i}\tilde{m}}\left(z\right)\right|^2
        -\left|\mathrm J_{\frac12+\mathrm{i}\tilde{m}}\left(z\right)\right|^2\right],  \label{barpsipsi}\\[2mm]
    p_\psi&=\frac{\pi}{a^5}\frac{|\beta^+|^2+|\beta^-|^2}{\cosh\pi\tilde{m}}\int\mathrm d^3 k
    \frac{(-k\tau)k^2}{3\sqrt{(k/a)^2+(m+g\phi)^2}}
    \left[\left|\mathrm J_{-\frac12-\mathrm{i}\tilde{m}}\left(-k\tau\right)\right|^2+
    \left|\mathrm J_{\frac12+\mathrm{i}\tilde{m}}\left(-k\tau\right)\right|^2 \right]
    + \delta^3(\mathbf0)\text{ term}\nonumber\\
    &=\frac{4\pi^2 H^4}{\cosh\pi\tilde{m}}\int_0^1\mathrm{d}z\, \frac{z^5}{3\sqrt{z^2+\tilde m^2}}
    \left[\left|\mathrm J_{-\frac12-\mathrm{i}\tilde{m}}\left(z\right)\right|^2
        +\left|\mathrm J_{\frac12+\mathrm{i}\tilde{m}}\left(z\right)\right|^2\right] \nonumber\\
    &\qquad\qquad\qquad\qquad
    -\frac{2H^4}{a^3}\int\mathrm d^3k\, \delta^3(\mathbf 0)\frac{z^5}{3\sqrt{z^2+\tilde m^2}}\
    \left[\left|\mathrm J_{-\frac12-\mathrm{i}\tilde{m}}\left(-k\tau\right)\right|^2+
    \left|\mathrm J_{\frac12+\mathrm{i}\tilde{m}}\left(-k\tau\right)\right|^2 \right].  \label{p_psi}
\end{align}
In Eq. \eqref{barpsipsi}, we neglected the term associated with $\delta^3(\mathbf0)$.
For $\tilde m\ll1$, the equation of state is $\omega_\psi=p_\psi/\rho_\psi\approx1/3$, and for $\tilde m\gg1$, the equation of state becomes $\omega_\psi\to 0$.
It is interesting to note that the term including $\delta^3(\mathbf0)$, which arises from zero-point modes, has no impact on the state function of the Dirac field; therefore, it does not correspond to any physical observables.
Thus, in the following calculations, we will neglect the terms associated with $\delta^3(\mathbf 0)$.

\section{Fermion pair density \label{n_inf}}

The conserved current is defined by $ j^\mu=\bar{\psi}\gamma^\mu\psi$,
satisfying the current conservation $\partial_\mu j^\mu=0$.
It is associated with a conserved charge given by
\begin{align}
    \hat Q=\int\mathrm d^{3}x\hat j^{0}(x)=\int\mathrm d^{3}x\hat\psi^{\dagger}(x)\hat\psi(x).
\end{align}
We now point out the vanishing of this quantity by vacuum average.
Applying the fermion operators \eqref{quantized_2}, it arrives at the expression
\begin{align}
    \hat Q&=\frac{1}{(2\pi a)^3}\sum_{s=\pm}\int\mathrm d^3k
    \left[\hat{B}^\dagger_s(\mathbf{k})\hat{B}_s(\mathbf{k})\tilde{u}_s^{h=+\ \dagger}(\mathbf{k},\tau)\tilde{u}_s^{h=+}(\mathbf{k},\tau)
    -\hat{D}^\dagger_s(\mathbf{k})\hat{D}_s(\mathbf{k})\tilde{u}_s^{h=-\ \dagger}(\mathbf{k},\tau)\tilde{u}_s^{h=-}(\mathbf{k},\tau) \right] %\nonumber\\
    % &\qquad\qquad\qquad\qquad\qquad\qquad\qquad\qquad\qquad\qquad
    % +\frac{1}{a^3\mathrm{vol}(\mathbb{R}^3)}\sum_{s=\pm}\int\mathrm d^3k\delta^3(0)
    % \tilde{u}_s^{-\ \dagger}(\mathbf{k},\tau)\tilde{u}_s^{-}(\mathbf{k},\tau).
    \label{Q}
\end{align}
This operator is named the net fermion density, which is representing the number density of fermions minus the number density of anti-fermions.
Term related to the operator $\hat B^\dagger\hat B$ in Eq. \eqref{Q} represents the occupation number of fermions in phase space, while $\hat D^\dagger\hat D$ represents the occupation number of anti-fermions.

It is evident that, with the application of the eigenvectors \eqref{sol_psi_J}, we obtain a vanishing result for the conserved charge:
\begin{align}
    \left\langle Q\right\rangle = 0.
\end{align}
This result is also straightforward to verify by considring the in-vacuum state, that is, $\left\langle Q\right\rangle|_{\tau\to-\infty} = 0$.
Physically, this phenomenon occurs because generating a non-zero net fermion number requires a source of CP violation.

The total density of fermion pairs in comoving coordinate is defined as
\begin{align}
    \left\langle n\right\rangle=\frac1{\mathrm{vol}(\mathbb{R}^3)}\int\mathrm{d}^3x\left\langle:\hat{\psi}^\dagger(\mathbf x,\tau)\hat{\psi}(\mathbf x,\tau):\right\rangle, \label{n_x}
\end{align}
where $:\cdots:$ means the creation operators appear on the left of annihilation operators without anti-commutation.
This quantity means the number density of fermions plus the number density of anti-fermions,
which is analytically obtained by changing the minus sigh in Eq. \eqref{Q} into the plus sigh.
Applying the quantized Dirac field Eq.\eqref{quantized_2} and the vacuum-averaging expressions Eq.\eqref{average}, we derive the analytical expression for the fermion pair density:
\begin{align}
    \left\langle n\right\rangle =\frac{\pi}{a^3\cosh\pi\tilde{\xi}} \int\mathrm{d}^3k\,(- k\tau) & \left\{\left|\beta^+\right|^2\left[\left|\mathrm J_{\frac{1}{2}+\tilde{\eta}
        +\mathrm{i}\tilde{\xi}}\left(-k\tau\right)\right|^2
        +\left|\mathrm J_{-\frac{1}{2}+\tilde{\eta}+\mathrm{i}\tilde{\xi}}\left(-k\tau\right)\right|^2\right]\right. \nonumber \\
    & \left.+\left|\beta^-\right|^2\left[\left|\mathrm J_{-\frac{1}{2}-\tilde{\eta}-\mathrm{i}\tilde{\xi}}\left(-k\tau\right)\right|^2
        +\left|\mathrm J_{\frac{1}{2}-\tilde{\eta}-\mathrm{i}\tilde{\xi}}\left(-k\tau\right)\right|^2\right]\right\}, \label{n1}
\end{align}
Expanding the Bessel functions about parameter $\tilde\eta$ to its first order, it arrives
\begin{align}
    \left\langle n\right\rangle =\frac{4\pi^2}{\cosh\pi\tilde{\xi}} \int_0^1\mathrm{d}z\,z^3
        & \left\{\left(\left|\beta^+\right|^2+\left|\beta^-\right|^2\right)\left[\left|\mathrm J_{\frac{1}{2}+\mathrm{i}\tilde{\xi}}(z)\right|^2
        +\left|\mathrm J_{-\frac{1}{2}-\mathrm{i}\tilde{\xi}}(z)\right|^2\right]\right. \nonumber\\
        &+\tilde{\eta}\left(\left|\beta^+\right|^2-\left|\beta^-\right|^2\right)
            \left[\frac{\partial \mathrm J_\nu(z)}{\partial\nu}\right|_{\nu=\frac12+\mathrm{i}\tilde{\xi}}\mathrm J_{\frac12-\mathrm{i}\tilde{\xi}}(z)+ \mathrm{c.c.} \nonumber \\
        &\qquad\qquad\qquad\qquad\quad+\frac{\partial \mathrm J_\nu(z)}{\partial\nu}\Bigg|_{\nu=-\frac12+\mathrm{i}\tilde{\xi}}\mathrm J_{-\frac12-\mathrm{i}\tilde{\xi}}(z)+\mathrm{c.c.}\Bigg]\Bigg\}
            +\mathcal O\left(\tilde{\eta}^2\right), \label{n2}
\end{align}
with $z=-k\tau$ and c.c. denoting the complex conjugate of the previous term.
Here, $-k\tau<1$ represents the majority of the contribution, as plotted in \textcolor{blue}{Ref. \cite{production8}}.
Thus, the upper limit of the integral is set 1 rather than infinity to avoid divergence.
%This approach will also be used in Sec. \ref{Obv}.

Now, to calculate the integrals in Eq. \eqref{n2}, the definition of Bessel function and its derivative to the order are needed:
\begin{subequations}
\begin{align}
    J_\nu(z) & =\sum_{k=0}^\infty\frac{(-1)^k}{k!\Gamma(k+\nu+1)}\left(\frac z2\right)^{2k+\nu}, \\
    \frac{\mathrm{d}J_\nu(z)}{\mathrm{d}\nu} & =J_\nu(z)\ln(z/2)-\sum_{k=0}^\infty\frac{(-1)^k\psi_M(k+\nu+1)}{k!\Gamma(k+\nu+1)}\left(\frac z2\right)^{2k+\nu},
\end{align}\end{subequations}
where $\Gamma(z)$ and $\psi(z)$ are the gamma function and digamma function respectively. Then the fermion density becomes
\begin{align}
    \left\langle n\right\rangle=&\langle n_0\rangle (1-\ln4\cdot\tilde{\eta}\tanh\pi\tilde{\xi})-\langle n^\prime\rangle
    -\tilde{\eta}\left(4\pi^2H^3\right)\frac{\tanh\pi\tilde{\xi}}{\cosh\pi\tilde{\xi}}\cdot \nonumber\\
    &\mathrm{Re}\sum_{h=\pm}\sum_{k=0}^\infty\sum_{l=0}^\infty \frac{(-1)^k(-1)^l\psi_M(h/2+\mathrm{i}\tilde{\xi}+k+1)}
        {2^{2k+2l+h}k!l!(k+l+h/2+2)\Gamma(h/2+\mathrm{i}\tilde{\xi}+k+1)\Gamma(h/2-\mathrm{i}\tilde{\xi}+l+1)}
        +\mathcal O\left(\tilde{\eta}^2\right).  \label{n3}
\end{align}
%\twocolumn
Here, $\langle n_0\rangle$ means the result at $\tilde\eta=0$, given by
\begin{align}
    &\langle n_0\rangle=\langle n\rangle|_{\tilde\eta=0}\nonumber\\
    &\propto \frac{1}{\cosh\pi\tilde\xi}\int \mathrm d z\,z^3\left[\left|\mathrm J_{\frac12+\mathrm{i}\tilde{\xi}}\left(z\right)\right|^2
        +\left|\mathrm J_{-\frac12-\mathrm{i}\tilde{\xi}}\left(z\right)\right|^2\right] \nonumber\\
    &\propto \frac{1}{|\Gamma(1/2+\mathrm i \tilde\xi)|^2 \cosh\pi\tilde\xi}=1/\pi.
\end{align}
This implies that the fermion pair density $\langle n_0\rangle$ is independent of the parameter $\tilde\xi$, indicating insensitivity to both Yukawa interaction strength and inflationary models,
and remains a constant
\begin{align}
    \langle n_0\rangle=0.2122(4\pi^2 H^3),
\end{align}
which has also been numerically verified with minimal error.
The symbol $\langle n^\prime\rangle$ in Eq. \eqref{n3} is given by the exact numerical result
\begin{align}
    &\langle n^\prime\rangle
    =-\tilde{\eta}\left(2\pi H\right)^3\frac{\tanh\pi\tilde{\xi}}{\cosh\pi\tilde{\xi}} %\nonumber\\
    \cdot\int_0^1\mathrm{d}z\,z^3\ln z\left[\left|\mathrm J_{\frac{1}{2}+\mathrm{i}\tilde{\xi}}(z)\right|^2+\left|\mathrm J_{-\frac{1}{2}-\mathrm{i}\tilde{\xi}}(z)\right|^2\right] \nonumber\\
    &=0.1414(4\pi^2 H^3)\tilde\eta \tanh\pi\tilde{\xi}.   \label{n}
\end{align}

\section{Scalar perturbation induced by the fermionic field \label{scalar_pert}}

The scalar field driving cosmic inflation can be decomposed into its background component $\phi_0$ and its perturbation component $\delta\phi(\mathbf x,t)$, i.e., $\phi=\phi_0+\delta\phi$.
By varying the action Eq. (M1) with respect to $\phi$, we obtain the equation of motion for the perturbed field
\begin{align}
    \left(\frac{\partial^2}{\partial t^2}+3H\frac\partial{\partial t}-\frac{\nabla^2}{a^2}+V_{\phi\phi}\right)\delta\phi=g({\bar{\psi}} {\psi}-\langle{\bar{\psi}} {\psi}\rangle), \label{delta_phi_x}
\end{align}
where $V_{\phi\phi}=\mathrm d^2V/\mathrm d\phi^2$ represents the second derivative of the potential with respect to the scalar field.
The subtraction of the vacuum average on the right-hand side ensures that the perturbed field $\delta\phi$ averages to be zero ($\langle\delta\phi\rangle=0$).
Then, in momentum space, Eq. \eqref{delta_phi_x} becomes
\begin{align}
    \left(\frac{\mathrm{d}^2}{\mathrm{d}\tau^2}+k^2-\frac{a^{\prime\prime}}a+a^2V_{\phi\phi}\right)(a\delta\phi) %\nonumber\\
    =ga^3\left({\bar\psi}*{\psi}-\langle{\bar\psi}*{\psi}\rangle\right), \label{delta_phi_k}
\end{align}
where the prime symbol $'$ is the derivative with respect to conformal time $\tau$ and $*$ denotes the convolution:
\begin{align}
    f * g(\mathbf{k})=\int \mathrm{d}^3 k^{\prime} f\left(\mathbf{k}^{\prime}\right) g\left(\mathbf{k}-\mathbf{k}^{\prime}\right).
\end{align}
Since Eq. \eqref{delta_phi_k} involves to the first-order scalar perturbations, we neglect all terms beyond the leading order,
indicating the approximations $a\sim-1/H\tau$, $H\sim\text{constant}$, and $V_{\phi\phi}/H^2\sim0$.
As a result, the scalar perturbation operator is solved in term of an integral of Green's function
\begin{align}
    \delta & \phi(\mathbf{k},\tau)=u_\mathbf{k}(\tau)\hat{b_\mathbf{k}}+u_\mathbf{k}^*(\tau)\hat{b}_\mathbf{k}^\dagger+\frac ga\int\mathrm{d}\tau^{\prime}\mathrm{G}_k(\tau,\tau^{\prime}){a^3(\tau')}\cdot  \nonumber\\
    &\int\mathrm{d}^3p \left[\hat{\bar{\psi}}(\mathbf{k}-\mathbf{p},\tau')\hat{\psi}(\mathbf{p},\tau')
        -\left\langle\hat{\bar{\psi}}(\mathbf{k}-\mathbf{p},\tau')\hat{\psi}(\mathbf{p},\tau')\right\rangle\right]. \label{delta_phi_k_1}
\end{align}
In this equation, $u_\mathbf{k}$ represents the scalar perturbation reduced from vacuum fluctuations, and $\mathrm{G}_k(\tau,\tau^{\prime})$ is the Green's function, expressed as \textcolor{blue}{\cite{blue1,blue2}}
\begin{align}
    G_k(\tau,\tau^{\prime})=&\frac1{k^3\tau\tau^{\prime}}\left[(1+k^2\tau\tau^{\prime})\sin k(\tau-\tau^{\prime})%\right. \nonumber\\
      +k(\tau'-\tau)\cos k(\tau-\tau^{\prime})\right]\theta(\tau-\tau^{\prime}). \label{G}
\end{align}

The curvature perturbation is represented by the variable $\hat{\mathcal{R}}=(H/\dot{\phi})\delta\phi$
which consists of two components: one arising from vacuum fluctuations, corresponding to the term $u_\mathbf{k}$, and the other from fermion-sourced perturbation,
given by the integral term in Eq. \eqref{delta_phi_k_1}.
The correlation function of the sourced curvature perturbation is then expressed in the form of a Green's function integral:
\begin{align}
    \left\langle\hat{\mathcal{R}}^{(\mathrm s)}(\mathbf{k},\tau)\hat{\mathcal{R}}^{(\mathrm s)}(\mathbf{k}^{\prime},\tau)\right\rangle = \frac{g^2H^2}{a^2(\tau)\dot{\phi}^2}
        &\int\mathrm{d}\tau_1\,G_k(\tau,\tau_1)a^3(\tau_1)\int\mathrm{d}\tau_2\,G_k(\tau,\tau_2)a^3(\tau_2)\cdot   \nonumber\\
    &\int\mathrm{d}^3p \int\mathrm{d}^3p'
    \left[ \left\langle\hat{\bar\psi}(\mathbf{k}-\mathbf{p},\tau_1)\hat{\psi}(\mathbf{p},\tau_1)\hat{\bar\psi}(\mathbf{k}'-\mathbf{p}',\tau_2)\hat{\psi}(\mathbf{p}',\tau_2)\right\rangle- \right.  \nonumber\\
    &\quad\quad\quad\quad\quad\quad\ \left. \left\langle\hat{\bar\psi}(\mathbf{k}-\mathbf{p},\tau_1)\hat{\psi}(\mathbf{p},\tau_1)\right\rangle
        \left\langle\hat{\bar\psi}(\mathbf{k}'-\mathbf{p}',\tau_2)\hat{\psi}(\mathbf{p}',\tau_2)\right\rangle\right]. \label{Rs}
\end{align}
The vacuum expectation value on the right-hand side includes four field operators. By applying Wick's theorem \textcolor{blue}{\cite{Wick}}, it is then decomposed into
three terms, each involving the product of two vacuum expectation values,
\begin{align}
&\left\langle\hat{\bar\psi}(\mathbf{k}-\mathbf{p})\hat{\psi}(\mathbf{p})\hat{\bar\psi}(\mathbf{k}^{\prime}-\mathbf{p}^{\prime})\hat{\psi}(\mathbf{p}^{\prime})\right\rangle %\nonumber\\
=\left\langle\hat{\bar{\psi}}(\mathbf{k}-\mathbf{p})\hat{\psi}(\mathbf{p})\right\rangle\left\langle\hat{\bar{\psi}}(\mathbf{k}'-\mathbf{p}')\hat{\psi}(\mathbf{p}')\right\rangle \nonumber\\
&\quad-\left\langle\hat{\bar{\psi}}(\mathbf{k}-\mathbf{p})\hat{\bar{\psi}}(\mathbf{k}^{\prime}-\mathbf{p}^{\prime})\right\rangle\left\langle\hat{\psi}(\mathbf{p})\hat{\psi}(\mathbf{p}^{\prime})\right\rangle %\nonumber\\
    +\left\langle\hat{\bar{\psi}}(\mathbf{k}-\mathbf{p})\hat{\psi}(\mathbf{p}^{\prime})\right\rangle\left\langle\hat{\psi}(\mathbf{p})\hat{\bar{\psi}}(\mathbf{k}^{\prime}-\mathbf{p}^{\prime})\right\rangle,  \label{Wick}
\end{align}
which has accounted for the anti-symmetry when two fermion operators are exchanged.
Note that the first term on the right-hand side of Eq. \eqref{Wick} is simply the last product of the two vacuum expectation values from the right-hand side of Eq. \eqref{Rs},
and hence, it cancels out.
The second term vanishes directly because $\langle\hat\psi\hat\psi\rangle=0$.
Thus, only the third term is nonzero.

Using the expression for the quantized fermion field, the unequal time correlation between fermion operator and its charge conjugate is given by
\begin{align}
    \left\langle\hat{\bar{\psi}}(\mathbf{p},\tau_1)\hat{{\psi}}(\mathbf{p}',\tau_2) \right\rangle
    =& \frac{\pi\delta^3(\mathbf{p}-\mathbf{p}')}{\left[a(\tau_1)a(\tau_2)\right]^{3/2}}
    \frac{\left(p^2\tau_1\tau_2\right)^{1/2}}{\cosh\pi\tilde{m}}\left(|\beta^+|^2+|\beta^-|^2\right)\cdot  \nonumber\\
    &\left[\mathrm J_{ -\frac12-\mathrm i \tilde m}(-p\tau_1)\mathrm J_{ -\frac12+\mathrm i\tilde m}(-p\tau_2)
        -  \mathrm J_{\frac12-\mathrm i \tilde m}(-p\tau_1)\mathrm J_{\frac12+\mathrm i\tilde m}(-p\tau_2) \right].
\end{align}
In the super-horizon limit $-k\tau\to0$, the correlation function of sourced scalar perturbation \eqref{Rs} arrives at
\begin{align}
    &\left\langle\hat{\mathcal{R}}^{(s)}(\mathbf{k},\tau)\hat{\mathcal{R}}^{(s)}(\mathbf{k}^{\prime},\tau)\right\rangle
    =\pi^2 g^2 \delta^3(\mathbf{k}-\mathbf{k}^{\prime}) \Bigg(\frac{H^2}{\dot{\phi}}\Bigg)^2 \left(\frac{1}{\cosh\pi\tilde{m}}\right)^2\frac{1}{k^3}\cdot
    \int_0^\infty\mathrm{d}x\int_0^1\mathrm{d}z_1\int_0^1\mathrm{d}z_2 \nonumber\\
    &\quad\quad x^4(\sin z_1-z_1\cos z_1)(\sin z_2-z_2\cos z_2) \left|\mathrm J_{-\frac12-\mathrm i \tilde m}(xz_1)\mathrm J_{-\frac12+\mathrm i\tilde m}(xz_2)-
    \mathrm J_{\frac12-\mathrm i \tilde m}(xz_1)\mathrm J_{\frac12+\mathrm i\tilde m}(xz_2)
        \right|^2, \label{corr_R}
\end{align}
where the integral variables $z_1=-k\tau_1$, $z_2=-k\tau_2$, and $x=p/k$ are introduced.

\section{Tensor perturbation induced by the fermionic field  \label{tensor_pert}}

The equation of motion of tensor perturbation is derived from the Einstein equation and is written as
\begin{align}
    \left(\frac{\partial^2}{\partial\tau^2}+k^2-\frac{a^{\prime\prime}}a\right)(ah_{ij})=\frac{2a^3}{M_p^2}\Pi_{ij,mn}T^{mn}(\mathbf{k},\tau),
\end{align}
where $M_p=(8\pi G)^{-1/2}$ is the Planck mass and $\Pi_{ij,mn}(\mathbf{k})=P_{im}(\mathbf{k})P_{jn}(\mathbf{k})-\frac12P_{ij}(\mathbf{k})P_{mn}(\mathbf{k})$ is the transverse-traceless operator,
with the projection operator $P_{ij}(\mathbf{k})=\delta_{ij}-k_i k_j/k^2$. The energy-momentum tensor for the fermion field is obtained by varying the Lagrangian with respect to the metric \textcolor{blue}{\cite{Dirac1,Dirac2}}:
\begin{align}
&T^{mn}(\mathbf x,\tau) =\frac2{\sqrt{-g}}\frac{\delta\mathcal{L}}{\delta g_{mn}} \nonumber\\
&=\frac{\dot{1}}{4}\bar{\psi}\left[\gamma^m\overrightarrow{D}^n + \overleftarrow{D}^n\gamma^m\right]\psi+\frac12g^{mn}\mathcal{L}+(m\leftrightarrow n), \label{T_x}
\end{align}
where $\bar\psi\overleftarrow{D}^n=-(\partial_\mu\bar\psi-\bar\psi\Gamma_\mu)$.
After Fourier transform, the fermionic energy-momentum tensor in momentum space becomes
\begin{align}
    &T_{mn}(\mathbf{k},\tau)= \nonumber\\
    &\frac{\mathrm i}{4}\int{\mathrm{d}^3p}\, \bar{\psi}(\mathbf{k}-\mathbf{p},\tau) %\nonumber\\
        \left[\gamma^mp^n-\gamma^n(k-p)^m\right]\psi(\mathbf{p})\nonumber\\
    &+\frac{\mathrm i \dot{a}}{4}\int{\mathrm{d}^3p}\, \bar{\psi}(\mathbf{k}-\mathbf{p},\tau)[\gamma^m\gamma^0\gamma^n+\gamma^0\gamma^n\gamma^m]\psi(\mathbf{p}) \nonumber\\
    &+(m\leftrightarrow n). \label{T_k}
\end{align}
Here, we drop the term $g^{mn}\mathcal{L}$ in Eq. \eqref{T_x} due to the traceless condition $\Pi_{ij,mn}g^{mn}=0$.

We then decompose the tensor perturbation $h_{ij}$ into the polarized states:
\begin{align}
    h_{ij}(\mathbf{k},\tau)=h_+(\mathbf{k},\tau)e_{ij}^+(\mathbf{k})+h_-(\mathbf{k},\tau)e_{ij}^-(\mathbf{k}),
\end{align}
where $e_{ij}^\pm$ is the polarization tensor, satisfying the contraction conditions $e_{ij}^\lambda({\mathbf{k}})e_{ij}^{\lambda'}({\mathbf{k}})^{\dagger}=\delta_{\lambda\lambda'}$, $e_{ij}^{\lambda}({\mathbf{k}})k^i=0$,
and $e_{ij}^\pm({\hat{\mathbf{k}}})\Pi_{mn}^{ij}({\mathbf{k}})=e_{mn}^\pm({\mathbf{k}})$.
The polarization tensors $e_{ij}^\pm$ are constructed by the polarization vectors $e_{ij}^\pm=\frac1{\sqrt{2}}\epsilon_i^{(\pm)}\otimes\epsilon_j^{(\pm)}$.
Using the Green's function method, the solution for the tensor mode $\hat h_\lambda^{(\mathrm s)}$ is expressed as
\begin{align}
    &\hat{h}_\lambda(\mathbf{k},\tau)=v_\mathbf{k}\left(\tau\right)\hat{c}_\mathbf{k}+v_\mathbf{k}^*\left(\tau\right)\hat{c}_\mathbf{k}^\dagger+\frac2{M_p^2 a(\tau)}e_{mn}^\lambda(\mathbf{k})^\dagger\cdot \nonumber\\
    &\int\mathrm{d}\tau^{\prime}G_{\mathbf{k}}(\tau,\tau^{\prime})a^3(\tau^{\prime})
        \int\operatorname{d}^3p\, \frac{\mathrm i}{4}\hat{\bar{\psi}}(\mathbf{k}-\mathbf{p},\tau^{\prime}) \nonumber\\
    &\left[\gamma^mp^n-\gamma^n(k-p)^m+\frac{\dot{a}}{2}(\gamma^m\gamma^0\gamma^n+\gamma^0\gamma^n\gamma^m)\right]\hat{\psi}(\mathbf{p},\tau) \nonumber\\
    &+(m\leftrightarrow n),   \label{h}
\end{align}
where $v_\mathbf{k}\left(\tau\right)$ represents the tensor perturbations from vacuum fluctuations and the Green's function $G_{\mathbf{k}}(\tau,\tau^{\prime})$ is shown in Eq. \eqref{G}.

The contraction over the polarization tensor and the gamma matrix gives $e_{ij}^\pm(\hat{\mathbf{k}})\gamma^ip^j=e_{ij}^\pm(\hat{\mathbf{k}})\gamma^jp^i=p^\pm\gamma^\pm/\sqrt{2}$, with $\gamma^\pm=\gamma^1\pm\mathrm i \gamma^2$ and $p^\pm=p^1\pm\mathrm i p^2$.
The vacuum expectation about the contracted polarization tensor $e_{ij}^\pm\gamma^ip^j$ and the quantized fermion field $\hat\psi$ is given by
\begin{align}
    \left\langle\hat{\bar{\psi}}(\mathbf{k}^{\prime}-\mathbf{p}^{\prime},\tau_1)\mathrm{e}_{ij}^\pm(\mathbf{k})\gamma^ip^j\psi(\mathbf{p},\tau_2)\right\rangle %\nonumber\\
    =&-\frac{\mathrm i\pi}{2\sqrt{2}}\frac{p_\pm^2}p\frac{\delta^3(\mathbf{k'-p'-p})}{\left[a(\tau_1)a(\tau_2)\right]^{3/2}}\frac{|\beta^+|^2-|\beta^-|^2}{\cosh\pi\tilde{m}}(p^2\tau_1\tau_2)^{1/2}\cdot \nonumber\\
    &\quad \left[\mathrm J_{-\frac12-\mathrm{i}\tilde{m}}(-p\tau_1)\mathrm J_{\frac12+\mathrm{i}\tilde{m}}(-p\tau_2)
    -\mathrm J_{\frac12-\mathrm{i}\tilde{m}}(-p\tau_1)\mathrm J_{-\frac12+\mathrm{i}\tilde{m}}(-p\tau_2)\right].
\end{align}
In addition, the contribution from the remaining gamma matrices vanishes, i.e.,
$\gamma^m\gamma^0\gamma^n+\gamma^0\gamma^n\gamma^m+(m\leftrightarrow n)=0$, because $\{\gamma^m,\gamma^0\}=0$.
Finally, we have the correlation function for the sourced tensor perturbation
\begin{align} \label{corr_h}
    &\left\langle\hat{h}_\lambda^{(\mathrm s)}(\mathbf{k},\tau)\hat{h}_\lambda^{(\mathrm s)\dagger}(\mathbf{k}',\tau)\right\rangle
        =\frac{\pi^2H^4}{2M_p^4a^2(\tau)}\left(\frac{\tanh\pi\tilde{ m}}{\cosh\pi\tilde{ m}}\right)^2\delta^3(\mathbf{k}-\mathbf{k}')
    \int\mathrm{d}\tau_1G_k(\tau,\tau_1)\int\mathrm{d}\tau_2G_k(\tau,\tau_2)\cdot \nonumber\\
    &\quad\quad\quad\quad \int\mathrm d^3p\, p_+^2p_-^2\tau_1\tau_2\ \left|\mathrm J_{-\frac12-\mathrm{i}\tilde{ m}}(-p\tau_1)\mathrm J_{\frac12+\mathrm{i}\tilde{ m}}(-p\tau_2)
    -\mathrm J_{\frac12-\mathrm{i}\tilde{ m}}(-p\tau_1)\mathrm J_{-\frac12+\mathrm{i}\tilde{ m}}(-p\tau_2) \right|^2 \nonumber\\
    &\approx\frac{16\pi^3}{15}\frac{H^4}{M_p^4} \left(\frac{\tanh\pi\tilde{ m}}{\cosh\pi\tilde{ m}}\right)^2 \frac{1}{k^3}\delta^3(\mathbf{k}-\mathbf{k}')
        \int_0^\infty\mathrm{d}x\int_0^1\mathrm{d}z_1\int_0^1\mathrm{d}z_2(\sin z_1-z_1\cos z_1)(\sin z_2-z_2\cos z_2)\cdot \nonumber\\
    &\quad\quad\quad\quad x^4\left|\mathrm J_{-\frac12-\mathrm{i}\tilde{ m}}(xz_1)\mathrm J_{\frac12+\mathrm{i}\tilde{ m}}(xz_2)-
    \mathrm J_{\frac12-\mathrm{i}\tilde{ m}}(xz_1)\mathrm J_{-\frac12+\mathrm{i}\tilde{ m}}(xz_2)
        \right|^2,
\end{align}
with $z_1=-k\tau_1$, $z_2=-k\tau_2$, and $x=p/k$.
In the above equations, the super-horizon limit $k\tau\to0$ and the integral about the polarized momentum $\int\mathrm d\Omega\ \hat p_+^2\hat p_-^2=32\pi/15$ (with $\hat p_+\hat p_-=\hat p_x^2+\hat p_y^2=\sin^2\theta$) are applied.

%\appendix
\section{Solutions of the Dirac-like equations in terms of Bessel functions\label{Bessel} }

Fermions that propagate to the asymptotic future, denoted as $\tilde{u}$, represent particles with chiral state $\mu_+ = +\mu$,  where $\mu$ is defined in Eq. (M13).
While fermions propagating to the asymptotic past $\tilde v$ is about the state $\mu_-=-\mu$.
They are analytically expressed in terms of Bessel functions:
%\onecolumn
\begin{subequations} \label{sol_psi_J}
\begin{align}
    &\tilde u_{s=+}^{h=+}/\tilde v_{s=+}^{h=+}(\mathbf{k},\tau) =  \Bigg(\frac{-\pi k\tau}{2\cosh{\pi\tilde{\xi}}}\Bigg)^{1/2}
        \begin{pmatrix}\begin{pmatrix}1\\0\end{pmatrix}\mathrm J_{\pm(-\frac12+\tilde{\eta}+\mathrm{i}\tilde{\xi})}(-k\tau)\\[1mm]
        \mp\frac{\mathrm{i}}k\begin{pmatrix}{k_z}\\{k_+}\end{pmatrix}\mathrm J_{\pm(\frac12+\tilde{\eta}+\mathrm{i}\tilde{\xi})}(-k\tau)\end{pmatrix}, \\
    &\tilde u_{s=-}^{h=+}/\tilde v_{s=-}^{h=+}(\mathbf{k},\tau) =  \Bigg(\frac{-\pi k\tau}{2\cosh{\pi\tilde{\xi}}}\Bigg)^{1/2}
        \begin{pmatrix}\begin{pmatrix}0\\1\end{pmatrix}\mathrm J_{\pm(-\frac12+\tilde{\eta}+\mathrm{i}\tilde{\xi})}(-k\tau)\\[1mm]
     \mp\frac{\mathrm{i}}{k}\begin{pmatrix}{k_-}\\{-k_z}\end{pmatrix}\mathrm J_{\pm(\frac12+\tilde{\eta}+\mathrm{i}\tilde{\xi})}(-k\tau)\end{pmatrix},  \\
    &\tilde u_{s=+}^{h=-}/\tilde v_{s=+}^{h=-}(\mathbf{k},\tau) =  \Bigg(\frac{-\pi k\tau}{2\cosh{\pi\tilde{\xi}}}\Bigg)^{1/2}
        \begin{pmatrix}\begin{pmatrix}1\\0\end{pmatrix}\mathrm J_{\pm(\frac12-\tilde{\eta}-\mathrm{i}\tilde{\xi})}(-k\tau)\\[1mm]
    \pm\frac{\mathrm{i}}{k}\begin{pmatrix}{k_z}\\{k_+}\end{pmatrix}\mathrm J_{\pm(-\frac12-\tilde{\eta}-\mathrm{i}\tilde{\xi})}(-k\tau)\end{pmatrix},  \\
    &\tilde u_{s=-}^{h=-}/\tilde v_{s=-}^{h=-}(\mathbf{k},\tau) =  \Bigg(\frac{-\pi k\tau}{2\cosh{\pi\tilde{\xi}}}\Bigg)^{1/2}
        \begin{pmatrix}\begin{pmatrix}0\\1\end{pmatrix}\mathrm J_{\pm(\frac12-\tilde{\eta}-\mathrm{i}\tilde{\xi})}(-k\tau)\\[1mm]
    \pm\frac{\mathrm{i}}{k}\begin{pmatrix}{k_-}\\{-k_z}\end{pmatrix}\mathrm J_{\pm(-\frac12-\tilde{\eta}-\mathrm{i}\tilde{\xi})}(-k\tau)\end{pmatrix}.
\end{align}
\end{subequations}
%\twocolumn
%%\end{thebibliography}
%\begin{thebibliography}{References}
%
%\bibitem{chiral} Z. K. Tao, L. Ma, D. Li, \textit{et al}. Chiral anomaly from Dirac particles in the presence of a conformal uniform background magnetic field in de Sitter spacetime, Phys. Rev. D \textbf{109}, 043503 (2024).
%
%\bibitem{production8} D. J. H. Chung, L. L. Everett, H. Yoo, \textit{et. al.}, Gravitational fermion production in inflationary cosmology, Phys. Lett. B, \textbf{712(3)}, 147-154 (2012).
%
%\bibitem{blue1} X. B. Li, H. Wang, and J. Y. Zhu, Gravitational waves from warm inflation, Phys. Rev. D \textbf{97}, 063516 (2018).
%
%\bibitem{blue2} M. M. Anber, E. Sabancilar, Chiral gravitational waves from chiral fermions. Phys. Rev. D \textbf{96}, 023501 (2017).
%
%\bibitem{Wick} G. D. Mahan, \textit{Many-Particle Physics} (Springer Science \& Business Media, New York, 2000).
%
%\bibitem{Dirac1} M. O. Ribas, F. P. Devecchi, G. M. Kremer, Fermionic cosmologies with Yukawa-type interactions, Europhysics Letters, \textbf{93}, 19002 (2011).
%
%\bibitem{Dirac2} M. O. Ribas, F. P. Devecchi, G. M. Kremer, Cosmological model with fermion and tachyon fields interacting via Yukawa-type potential, Modern Physics Letters A, \textbf{31(06)}, 1650039 (2016).
%
%\end{thebibliography}

\end{widetext}

\end{document}